\documentclass[acmsmall]{acmart}

\makeatletter
\@ifpackageloaded{amssymb}{}{}
\makeatother

\usepackage{adjustbox}
\usepackage{amsmath}
\usepackage{amssymb}
\usepackage{booktabs}
\usepackage{enumitem}
\usepackage{listings}
\usepackage{multirow}
\usepackage{siunitx}
\usepackage{tabularx}
\usepackage{xcolor}
\usepackage{colortbl}
\usepackage{tcolorbox}
\tcbuselibrary{breakable}
\tcbuselibrary{skins}

\definecolor{gray}{RGB}{200,200,200}
\newtcolorbox{examplebox}[1]{
    colback=white,
    colframe=gray!50,
    boxrule=1pt,
    arc=3pt,
    title=#1,
    fonttitle=\bfseries,
    enhanced,
    breakable,
    coltitle=black,
    before upper={\parindent0pt}
}

\definecolor{gray}{RGB}{200,200,200}
\newtcolorbox{snippetbox}[1][]{%
    enhanced,
    colback=white,
    colframe=gray!75!black,
    boxrule=0.5pt,
    arc=2pt,
    left=3mm,
    right=3mm,
    top=2mm,
    bottom=2mm,
    boxsep=1mm,
    fontupper=\small,
    #1
}

\usepackage{fancyvrb}

\newenvironment{codeverbatim}
 {\VerbatimEnvironment
  \begin{adjustbox}{margin=10pt,center}
  \begin{minipage}{0.9\textwidth}
  \begin{BVerbatim}[formatcom=\small]}
 {\end{BVerbatim}
  \end{minipage}
  \end{adjustbox}}

\newcommand{\header}[1]{\noindentparagraph{\emph{\textbf{#1}}}}

\newcommand{\rqi}{How accurately can LLMs identify trace links between documentation segments and their corresponding code elements?}
\newcommand{\rqii}{How effectively can LLMs explain the nature of relationships between documentation and code elements?}
\newcommand{\rqiii}{How completely can LLMs identify intermediate elements in documentation-to-code trace chains?}

\begin{document}

\title{Evaluating the Use of LLMs for Documentation to Code Traceability}

\author{Ebube Alor}
\affiliation{%
  \department{Data-driven Analysis of Software (DAS) Lab}
  \department{Department of Computer Science \& Software Engineering}
  \institution{Concordia University}
  \city{Montréal}
  \state{QC}
  \country{Canada}}
\email{ebubechukwu.alor@mail.concordia.ca}

\author{SayedHassan Khatoonabadi}
\affiliation{%
  \department{Data-driven Analysis of Software (DAS) Lab}
  \department{Department of Computer Science \& Software Engineering}
  \institution{Concordia University}
  \city{Montréal}
  \state{QC}
  \country{Canada}}
\email{sayedhassan.khatoonabadi@mail.concordia.ca}

\author{Emad Shihab}
\affiliation{%
  \department{Data-driven Analysis of Software (DAS) Lab}
  \department{Department of Computer Science \& Software Engineering}
 \institution{Concordia University}
 \city{Montréal}
 \state{QC}
 \country{Canada}}
\email{emad.shihab@concordia.ca}

\renewcommand{\shortauthors}{Alor et al.}

\begin{abstract}
    Large Language Models (LLMs) offer new potential for automating documentation-to-code traceability, yet their capabilities remain underexplored. We present a comprehensive evaluation of LLMs (Claude 3.5 Sonnet, GPT-4o, and o3-mini) in establishing trace links between various software documentation (including API references and user guides) and source code. We create two novel datasets from two open-source projects (Unity Catalog and Crawl4AI). Through systematic experiments, we assess three key capabilities: (1) trace link identification accuracy, (2) relationship explanation quality, and (3) multi-step chain reconstruction. Results show that the best-performing LLM achieves F1-scores of 79.4\% and 80.4\% across the two datasets, substantially outperforming our baselines (TF-IDF, BM25, and CodeBERT). While fully correct relationship explanations range from 42.9\% to 71.1\%, partial accuracy exceeds 97\%, indicating that fundamental connections are rarely missed. For multi-step chains, LLMs maintain high endpoint accuracy but vary in capturing precise intermediate links. Error analysis reveals that many false positives stem from naming-based assumptions, phantom links, or overgeneralization of architectural patterns. We demonstrate that task-framing, such as a one-to-many matching strategy, is critical for performance. These findings position LLMs as powerful assistants for trace discovery, but their limitations could necessitate human-in-the-loop tool design and highlight specific error patterns for future research.
\end{abstract}




\maketitle

\section{Introduction}
\label{sec:introduction}
Traceability in software engineering refers to the ability to establish and maintain relationships between various software artifacts produced throughout the software development lifecycle~\cite{cleland2014software, cleland2012software}. These artifacts include requirements, design documents, source code, test cases, and documentation~\cite{pfeiffer2020constitutes, ma2018automatic}. Effective traceability enables stakeholders to understand how different artifacts influence one another, facilitating tasks such as impact analysis, change management, and compliance verification~\cite{cleland2014software}. Documentation-to-code traceability focuses on linking natural language documentation to the corresponding code implementations~\cite{antoniol2002recovering, chen2011improving}. Documentation plays a vital role in software development by aiding in communication among team members, providing guidance for software usage, and serving as a reference for maintenance and future development~\cite{parnas2010precise}. It helps developers and users understand the system's functionality, design decisions, and usage instructions, which is essential for the success and longevity of software projects.

However, establishing accurate traceability between documentation and code presents significant challenges. Documentation is often written in free-flowing, natural language that is unstructured and may contain ambiguities~\cite{chen2011improving}. Traditional traceability methods, such as Information Retrieval (IR) techniques, rely on keyword matching and statistical analysis, which may not capture the deeper semantic relationships between documentation and code~\cite{de2006can, lucia2007recovering}. These methods suffer from low precision or recall, leading to incomplete or incorrect traceability links~\cite{de2006can, lucia2007recovering}. As software systems evolve and codebases grow in size and complexity, scalability becomes a critical concern~\cite{zimmermann2004preprocessing, gotel2012grand}. Existing methods may require substantial manual effort to maintain traceability links and may not adapt well to changes in the code or documentation~\cite{ruiz2023don, tian2021impact}. Additionally, they may be limited to specific programming languages or domains, hindering their applicability in diverse settings~\cite{gotel2012grand}.

Recent advancements in Large Language Models (LLMs) offer promising opportunities to address these challenges \cite{zheng2025towards, nam2024using, hou2024large, fan2023large}. LLMs, such as OpenAI's GPT series, have demonstrated exceptional capabilities in understanding and generating human-like text, as well as in processing and generating code~\cite{ozkaya2023application, nam2024using, chen2021evaluating, fan2023large}. Their ability to comprehend natural language semantics and code syntax suggests the potential for improving documentation-to-code traceability by bridging the gap between unstructured documentation and structured code artifacts. Despite the potential of LLMs, there is a lack of comprehensive studies evaluating their effectiveness in documentation-to-code traceability. 

To address this knowledge gap, we conduct a comprehensive evaluation of LLMs for documentation-to-code traceability tasks. We create two novel datasets from two modern open-source projects, both developed after the most recent LLMs training cutoff date (April 2024). These projects feature diverse documentation styles, including API references and tutorial-style documentation, along with their corresponding code artifacts at different granularity levels (classes, methods, and statements). Our methodology employs a systematic evaluation approach where we assess three leading LLMs (Claude Sonnet 3.5, GPT-4o, and o3-mini) against established baselines (TF-IDF, BM25, and CodeBERT) using a one-to-many matching strategy---where each documentation segment is evaluated against all potential code artifacts simultaneously rather than examining each possible pair individually. For each LLM, we examine its ability to establish trace links, explain relationships, and reconstruct dependency chains between documentation and code. Based on this, our study aims to investigate the following three research questions:

\header{RQ1:} \emph{\rqi}
While LLMs demonstrate code-text understanding, their fundamental capability to map documentation to precise code elements remains unproven. Traditional IR methods struggle with the documentation's hybrid structure (mix of formal specs and natural language) and implicit relationships. Establishing this baseline capability is critical before exploring advanced traceability applications. We define TF-IDF, BM25, and CodeBERT as our baselines. We find that the best-performing LLM achieves F1-scores of 79.4\% and 80.4\% on the Crawl4AI and Unity Catalog datasets, respectively, substantially outperforming our baselines (whose best F1-scores were 54.2\% and 69.3\%, respectively).

\header{RQ2:} \emph{\rqii}
Traceability can require more than binary links (is related vs not related). For example, maintenance tasks demand an understanding of how code implements documented functionality. While LLMs are capable of generating natural language explanations, their capacity to produce technically precise relationship descriptions (e.g., "implements validation rule" vs "configures service") remains unexplored. We evaluate whether LLMs advance beyond linkage to semantic understanding crucial for developer trust and actionable insights. We find that while fully correct relationship explanations range from 42.9\% to 71.1\%, partial accuracy exceeds 97\%, indicating that fundamental connections are rarely missed.

\header{RQ3:} \emph{\rqiii}
Real-world traceability often involves multi-step connections (e.g., documentation $\rightarrow$ interface $\rightarrow$ class $\rightarrow$ method). Prior work often focuses on direct links (documentation $\rightarrow$ method), but software evolution requires understanding dependency chains. We evaluate whether LLMs can reconstruct these pathways---a capability essential for impact analysis and change management in complex systems. We find that for multi-step chains, LLMs maintain high endpoint accuracy (below 2\% incorrect) but vary in capturing precise intermediate links from 13\% to 80\%. 

\vspace{\baselineskip}
While our results show that LLMs outperform traditional baselines in identifying trace links, they also reveal limitations, particularly in generating fully complete explanations and tracing multi-step dependencies. We then turn our analysis from what the LLMs achieve to how task-framing influences the performance. Specifically, we investigate the impact of context management strategies and additional contextual information on traceability performance. Based on our findings, we offer practical recommendations: for developers, LLMs can be valuable traceability assistants when paired with human oversight; and for tool designers, strategies like one-to-many linking and human-in-the-loop review can enhance usability. Finally, we outline directions for future LLM research to address current limitations in explanation depth and intermediate trace reconstruction.

\header{Contributions.} Our key contributions include:
\begin{itemize}
    \item \textbf{Dataset Creation:} We created two novel datasets specifically designed for evaluating documentation-to-code traceability using LLMs, which include varied documentation and code samples at different levels of granularity.
    \item \textbf{Comprehensive Evaluation:} We provide empirical evidence on LLMs for documentation-to-code traceability tasks, examining their ability to classify and map documentation to code accurately.
    \item \textbf{Relationship Identification:} We explore the capability of LLMs to identify and describe the specific types of relationships between documentation and code, moving beyond simple linkage to semantic understanding.
    \item \textbf{Practical Implications:} We discuss recommendations for integrating LLMs into software development practices to enhance traceability and support maintenance activities.
\end{itemize}

\header{Paper Organization.} The remainder of this paper is organized as follows. Section~\ref{sec:background} provides background information and reviews related work on documentation-to-code traceability and the application of LLMs in software engineering. Section~\ref{sec:methodology} describes the methodology used for our evaluation, including dataset preparation and experimental design. Sections~\ref{sec:rq1}, ~\ref{sec:rq2} and ~\ref{sec:rq3} presents the experimental results and analysis for each research question. Section~\ref{sec:discussion} discusses additional analysis in our paper. Section~\ref{sec:implications} discusses the implications of our findings for practitioners, tool designers, and researchers. Section ~\ref{sec:threats} highlights threats to the validity of the study. Finally, Section~\ref{sec:conclusion} concludes the paper.

\section{Background and Related Work}
\label{sec:background}
In this section, we provide an overview of the key concepts and terminology essential for understanding our work. First, we discuss software documentation and its complexities. Next, we look into software traceability, focusing specifically on the critical challenge of linking documentation to code. We then review traditional automated approaches for traceability and their limitations. Finally, we discuss the potential of LLMs in software engineering, particularly their promise for addressing documentation-to-code traceability challenges.

\subsection{Software Documentation}

Software documentation serves as the primary medium for communicating software system knowledge between developers, maintainers, and users \cite{forward2002relevance, ding2014knowledge}. While source code defines system behavior, documentation explains the intentions, usage patterns, and architectural decisions that shape a software system \cite{parnas2010precise}. This critical role makes documentation essential for both software use and maintenance, and a key target for traceability efforts.

Modern software documentation comprises multiple forms, each addressing distinct objectives and audiences \cite{aghajani2020software}. \emph{User documentation} targets end-users by presenting tutorials, usage guides, and troubleshooting steps. \emph{Technical documentation} details architectural decisions and design rationale \cite{ding2014knowledge}. At a more code-centric level, \emph{API documentation} often incorporates both automatically generated references (e.g., function signatures, endpoints) and explanatory text contextualizing system components \cite{maalej2013patterns}. These forms illustrate documentation's range from high-level guidance to deep technical specifics \cite{maalej2013patterns}.

However, maintaining documentation quality presents significant hurdles. A primary challenge is \emph{consistency}: documentation often becomes out-of-sync with evolving code, a phenomenon known as \emph{documentation decay} \cite{steidl2013quality, tan2024detecting, silva2023towards, aghajani2019software}. This risk is amplified in rapid development cycles \cite{silva2023towards}. While practices like \emph{documentation-as-code} \cite{cadavid2022documentation} and automated generation tools \cite{zhou2022docprompting, moreno2013jsummarizer, casas2021uses, khan2022automatic} help maintain structural references, they often fail to capture deeper \emph{semantic} linkages \cite{aghajani2019software}. Furthermore, documentation involves \emph{varying levels of abstraction}, from high-level concepts to implementation notes \cite{dvivedi2024comparative, maalej2013patterns}, and often relies on \emph{implicit} knowledge (like design rationale or constraints) not directly reflected in the code \cite{silva2023towards, uddin2015api}. These inherent complexities make establishing and maintaining clear connections to the codebase difficult \cite{tan2024detecting, uddin2015api, uddin2015api}.

In this work, we concentrate on two common and challenging documentation forms: user documentation (e.g., tutorials, guides) and API documentation \cite{silva2023towards}. These integrate natural language explanations with code references, providing an ideal testbed for evaluating LLM capabilities in linking textual and programmatic information \cite{maalej2013patterns}.

\subsection{Software Traceability and its Challenges}

Traceability in software engineering refers to the ability to connect related software artifacts---such as requirements, designs, code, and documentation---throughout the lifecycle \cite{cleland2014software, cleland2012software}. It is crucial for tasks like impact analysis, change management, and compliance verification, enabling stakeholders to understand how changes propagate and ensure software aligns with its goals \cite{cleland2014software}.

Historically, traceability links were maintained manually \cite{cleland2012software}. However, the scale of modern software systems, potentially involving millions of links, necessitates automation \cite{silva2023towards}. Beyond scale, software evolution poses a major traceability challenge: as features change and designs are updated, keeping artifacts synchronized becomes increasingly difficult \cite{cleland2014software, tan2024detecting}.

A particularly critical and challenging aspect is establishing reliable \emph{documentation-to-code traceability} \cite{antoniol2002recovering, uddin2015api}. This task is complicated by several factors inherent to the artifacts themselves:
\begin{itemize}
    \item \textbf{Abstraction Gap:} Documentation often uses natural language to describe high-level concepts, rationale, or usage scenarios \cite{lucia2007recovering}, while source code consists of formal, executable constructs \cite{antoniol2002recovering, marcus2003recovering}. A single documentation paragraph might relate to multiple, scattered code components \cite{Paper16, maalej2013patterns, uddin2015api}.
    \item \textbf{Evolution Mismatch:} Code is frequently refactored or renamed, while documentation updates may lag, leading to broken or outdated links \cite{Paper14, Paper19, aghajani2020software, tan2024detecting, silva2023towards}.
    \item \textbf{Format and Language Differences:} Documentation and code use different languages and structures, making direct mapping complex \cite{hayes2006advancing}.
    \item \textbf{Implicit Knowledge and Context:} Developers rely on implicit understanding of architecture or domain patterns to interpret documentation \cite{Paper17, uddin2015api}. Automated tools struggle to replicate this tacit knowledge, especially in large codebases or distributed systems (e.g., microservices) where context is fragmented \cite{guo2017semantically, zhou2015empirical, cleland2011traceability, nagappan2008influence}.
\end{itemize}
These factors significantly hinder the ability to establish and maintain accurate trace links between documentation and code using traditional methods \cite{antoniol2002recovering, uddin2015api}.

\subsection{Automated Traceability Approaches}
\label{sec:traceability-approaches}

To address the challenges of manual tracing, various automated methods for trace link discovery have been developed. Early efforts relied on \emph{information retrieval} (IR) techniques, such as keyword matching and text indexing \cite{antoniol2002recovering}. While useful for smaller projects, basic IR struggles with varying terminology and lacks deep contextual understanding \cite{hayes2006advancing, marcus2003recovering}.

Vector Space Models (VSM) \cite{lucia2007recovering, antoniol2002recovering} and techniques like \emph{Latent Semantic Indexing} (LSI) \cite{marcus2003recovering, hayes2006advancing} offered improvements by representing artifacts in vector spaces and capturing underlying semantic patterns, allowing identification of related concepts even with different terminology \cite{hayes2003improving, hayes2006advancing, Paper19}.

Subsequent \emph{machine learning} (ML) approaches utilized supervised techniques, training models on manually verified links to identify similar patterns \cite{cleland2010machine}. ML often augmented IR methods by incorporating deeper linguistic or domain-specific cues \cite{mills2018automatic, mills2017automating, marcen2020traceability}. Many modern tools combine IR, ML, and sometimes rule-based systems to handle the multifaceted nature of trace relationships \cite{chen2011improving}.

\emph{Natural language processing} (NLP) has become integral to many approaches, especially for interpreting documentation \cite{Paper16}. However, these automated methods still face limitations. IR-based techniques may miss semantic connections expressed with divergent vocabulary. ML approaches typically require significant labeled training data and may struggle with project-specific conventions or context shifts. Rule-based systems require ongoing maintenance as software evolves \cite{Paper20}. The persistent gap between the capabilities of these traditional methods and the complexity of accurately tracing documentation to code, particularly in capturing nuances like design intent or usage constraints \cite{maalej2013patterns}, drives the need for more robust, semantically enriched solutions \cite{Paper21, Paper22}.

\subsection{LLMs in Software Traceability}
LLMs, such as OpenAI's GPT-4o and Anthropic's Claude Sonnet 3.5 have demonstrated remarkable capabilities in natural language understanding and generation~\cite{karanikolas2023large, anthropic2024claude, hurst2024gpt}. Recent studies apply these models to software traceability and report superior performance over classical IR when linking requirements to code, mapping architecture documentation to source files, and tracing security requirements to goal models~\cite{ali2024establishing, fuchss2025enabling, hassine2024llm}.

This promise warrants a closer look at the traceability challenges that LLM properties may address. Traceability faces three persistent hurdles: the semantic gap between natural-language artifacts and source code, the scale of modern projects~\cite{ruiz2023don}, and the need for links that auditors can justify. First, state-of-the-art LLMs can accept context windows ranging from roughly 128\,000 to two million tokens, so a full requirements specification and a sizable code context can fit in a single prompt, easing the scale barrier~\cite{openai2025gpt41,anthropic2024claude, gemini2025pro}. Second, unified embeddings for text and code reduce the semantic gap and yield more precise candidate links than term-matching approaches~\cite{fuchss2025enabling, ali2024establishing}. Finally, chain-of-thought prompts and gradient-based token-highlight techniques let a model explain why two artifacts match, providing justifications stakeholders can inspect~\cite{north2024code, masoudifard2024leveraging}.

Despite these advances, LLM-based \emph{documentation-to-code} traceability remains largely unexplored. Developer documentation is markedly heterogeneous, ranging from README files and usage tutorials to API docs, which widens the semantic gap and introduces vocabulary drift~\cite{raglianti2023rise, raglianti2022topology, yang2025empirical}. Existing LLM studies concentrate on requirement-level links and rarely assess explanation fidelity or multi-hop chain recovery~\cite{north2024code,ali2024establishing}. Practitioners also caution that the opaque reasoning of current models complicates certification in audit-heavy domains~\cite{ruiz2023don}. We therefore perform a systematic investigation to clarify the strengths and limitations of LLMs before they can be trusted for documentation-centric traceability.

\section{Methodology}
\label{sec:methodology}
In this section, we describe the methodology employed to address our research questions. We discuss the criteria and process for selecting suitable LLMs for evaluation. We then explain how we create and structure our evaluation datasets, and conclude by describing the experimental design used to assess each LLM’s performance.

\subsection{LLM Selection}
\label{sec:llm_selection}

Evaluating the software traceability capabilities of LLMs requires selecting models that represent the current state-of-the-art while allowing for fair assessment on data they have not been trained on. Our selection process focused on identifying LLMs meeting several key criteria:

\begin{itemize}
    \item \textit{Strong Performance:} Demonstrates capabilities in complex reasoning and code-related tasks, essential for understanding documentation and source code relationships.
    \item \textit{Large Context Window:} Ability to process substantial amounts of text (documentation and code snippets) simultaneously, crucial for traceability, which often requires broad context.
    \item \textit{Cost-Effectiveness Features:} Mechanisms like API caching support to mitigate the costs associated with potentially large inputs and iterative analysis common in software engineering tasks.
    \item \textit{Recent and Verifiable Knowledge Cutoff:} A well-defined training data cutoff date. This is critical to ensure that we can subsequently create evaluation datasets (see Section~\ref{sec:project_selection}) guaranteed to be novel to the LLMs, thereby preventing data contamination and enabling a rigorous evaluation of their true traceability abilities on unseen projects.
\end{itemize}

Based on these criteria, we selected the following three LLMs:

\begin{itemize}
    \item \textbf{Claude 3.5 Sonnet (claude-3-5-sonnet-20240620):} Chosen for its state-of-the-art performance at the time of evaluation, particularly its reported strengths in reasoning and coding tasks \cite{anthropic2024claude}, combined with a large 200k token context window. It has an April 2024 knowledge cutoff.
    \item \textbf{OpenAI GPT-4o (gpt-4o-2024-08-06):} Selected as OpenAI's flagship model at the time, known for its strong general performance across various tasks, including code. It offers a 128k token context window and specific optimizations for efficiency \cite{openaimodelsgpt4o}. It has an October 2023 knowledge cutoff.
    \item \textbf{OpenAI o3-mini (o3-mini-2025-01-31):} We also later included OpenAI's o3-mini to evaluate an LLM specifically highlighted for advanced reasoning capabilities, a trait potentially beneficial for software engineering analysis, while also being designed for cost-efficiency \cite{openai2025o3, openaimodelso3mini}. It features a large 200k token context window. It also has an October 2023 knowledge cutoff like GPT-4o.
\end{itemize}

\subsection{Project Selection}
\label{sec:project_selection}

To evaluate the performance of LLMs in documentation-to-code traceability tasks, we required datasets containing software projects with well-documented codebases that were unlikely to be part of the selected LLMs' training data. Given that the latest of our selected LLMs has a knowledge cutoff date of April 2024, we focused on projects created after May~1, 2024, to minimize the chance of inclusion in the LLMs' training data. We also prioritized projects based on their popularity (as measured by GitHub star count) to ensure they were more recognized, while placing no constraints on programming language or application domain to allow diversity in the evaluation.

To do this, we used the SEART GitHub search tool \cite{seartgithubsearchweb, seartgithubsearchrepo}, which offers advanced filtering options not readily available in GitHub’s native search. We filtered projects created after May~1, 2024, and sorted them by descending star count, reflecting their popularity and likelihood of active maintenance. Starting with the most popular projects, we manually reviewed their documentation for clarity, comprehensiveness, and strong links to code elements, examining a total of 128 repositories. From this process, we selected the following top two projects that best met our criteria for detailed and high-quality documentation:

\begin{itemize}
    \item \textbf{Unity Catalog\footnote{\url{https://github.com/unitycatalog/unitycatalog}}:} An open-source, multimodal catalog that unifies governance across varied data and AI assets. Its repository features structured documentation on supported data formats, API usage, and security configurations---ideal for testing detailed, code-linked documentation.
    \item \textbf{Crawl4AI\footnote{\url{https://github.com/unclecode/crawl4ai}}:} A web crawler and scraper optimized for AI-centric tasks, offering features like concurrent crawling and metadata extraction. The project's documentation includes extensive usage examples and architecture notes, providing rich opportunities to evaluate how LLMs handle diverse documentation styles when tracing code elements.
\end{itemize}

\subsection{Dataset Creation}
\label{sec:dataset_creation}
\begin{figure}[t]
  \centering
  \includegraphics[width=\linewidth]{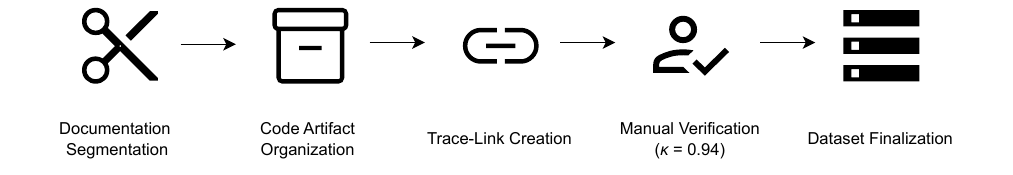}
  \caption{Five-step dataset-creation pipeline.}
  \label{fig:pipeline}
\end{figure}

To create our ground truth dataset, establishing verified links between documentation and code, we followed a five-step process (shown in Figure~\ref{fig:pipeline}). This process, detailed below, was designed to ensure the resulting dataset accurately reflects the relationships within the selected software projects and provides a reliable basis for evaluating LLM traceability performance. We illustrate these steps using a simplified, running example based on configuring a web crawler component, similar in nature to elements found in our selected project Crawl4AI.

% --- Running Example Introduction ---
\vspace{\baselineskip}
\noindent\textbf{Running Example Scenario:}

\begin{figure}[htbp]
  \centering
  \begin{snippetbox}
  \begin{lstlisting}[basicstyle=\small\ttfamily, breaklines=true, columns=flexible]
    ### Setting Concurrency Level
    
    To control the number of simultaneous requests the crawler
    makes, use the `max_concurrent_requests` setting in your
    Crawler configuration dictionary. A higher value increases
    speed but also resource usage. The default is 5.
    
    Example Usage:
    ```python
    crawler_settings = {
        "start_url": "https://example.com",
        "max_concurrent_requests": 10
    }
    my_crawler = Crawler(config=crawler_settings)
    
    ---------------------
    
    ### Newsletter Signup
    
    Stay informed about new features by signing up on our website.
    This step is optional and does not affect crawler settings.
\end{lstlisting}
\end{snippetbox}
  \caption{Illustrative documentation with two segments. The first segment (above the line) discusses concurrency settings; the second (below the line) briefly mentions a sign-up flow that does not map to any code artifact.}
  \label{fig:documentation-example}
\end{figure}

\noindent
\textit{Figure~\ref{fig:documentation-example} shows two consecutive segments from a hypothetical \texttt{docs/config.md} file. Segment A (``Setting Concurrency Level'') explicitly references how to configure the crawler, while Segment B (``Newsletter Signup'') is unrelated to the crawler's code and thus has no trace links. Figure~\ref{fig:code-example} presents the corresponding Python code snippet.}

\begin{figure}[htbp]
  \centering
  \begin{snippetbox}
  \begin{lstlisting}[language=python, basicstyle=\small\ttfamily, breaklines=true, columns=flexible]
    class Crawler:
        DEFAULT_CONCURRENCY = 5  # Default worker count
    
        def __init__(self, config):
            self.settings = config
            self.concurrency = self.settings.get(
                "max_concurrent_requests",
                Crawler.DEFAULT_CONCURRENCY
            )
            # ... other initialization ...
    
        def run(self):
            # uses self.concurrency to manage workers...
            pass
\end{lstlisting}
  \end{snippetbox}
  \caption{Illustrative Python code for the \texttt{Crawler} class, which is referenced by Segment A.}
  \label{fig:code-example}
\end{figure}

% --- End Running Example Introduction ---

\header{Step 1: Documentation Segmentation.}
The first step involved partitioning the project's documentation into manageable units, which we term \textbf{segments}. A segment is a semantically coherent portion of text, typically discussing a specific feature, API element, configuration option, or a distinct usage scenario. We identified segment boundaries often corresponding to section headings, logical breaks between topics, or distinct code examples. The goal was to create segments focused enough to allow for precise traceability to code, while retaining sufficient surrounding text to provide immediate context \cite{velasco2020automated}.

To achieve this, we divided each Markdown file into smaller segments based on top-level and subsection headings (e.g., \texttt{\#}, \texttt{\#\#}, \texttt{\#\#\#}) \cite{marcus2003recovering}. We used a script to detect these headings and extract the segments. We then manually checked each resulting segment for coherence. Segments that contained filler text or lacked any links to code (e.g., the ``Newsletter Signup'' in the running example) were marked as having zero links and excluded from the dataset.

\emph{Running Example:} In Figure~\ref{fig:documentation-example}, the content labeled ``Setting Concurrency Level'' was extracted as Segment~A. Because it explicitly mentions \texttt{max\_concurrent\_requests} and a crawler usage example, we preserved it for further analysis. Meanwhile, the ``Newsletter Signup'' text (Segment~B) was discarded as it did not discuss any relevant information.

\header{Step 2: Code Artifact Organization.}
Concurrently, we identified and cataloged relevant code elements, termed \textbf{code artifacts}. We established the \textbf{granularity} (level of detail) for our analysis by focusing on code artifacts typically involved in the software's architecture, feature-set, and public interface, as these are most likely to be referenced or described in documentation \cite{marcus2003recovering}. Specifically, we extracted classes, methods/functions, and class-level attributes/constants. We deliberately excluded finer-grained implementation details like individual statements within methods or local variables. This decision is based on the observation that documentation usually describes what components do or how to use them at an interface level, rather than detailing internal statement-by-statement logic; including such low-level details would likely add significant noise with few corresponding documentation links, a common consideration in traceability link recovery \cite{marcus2003recovering}. 

Each identified artifact was stored with its type (Class, Method, Function, Attribute, etc.), full path location, granularity levels (Class-level, Method-level, and Statement-level), and the source code content. Note that there is a subtle difference between artifact types and their granularities. For instance, a Python standalone function and a class method, are both considered to be Method-level. The same applies to a class attribute and a code statement, which we both consider to be Statement-level.

\emph{Running Example:} From the example code, we identify the following artifacts matching our granularity:
\begin{itemize}
% Use \texttt{} for inline code artifacts
\item \texttt{Crawler} (Type: Class, Location: \texttt{src/crawler.py}, Granularity: Class-level)
\item \texttt{Crawler.\_\_init\_\_} (Type: Method, Location: \texttt{src/crawler.py}, Granularity: Method-level)
\item \texttt{Crawler.DEFAULT\_CONCURRENCY} (Type: Attribute, Location: \texttt{src/crawler.py}, Granularity: Statement-level)
\end{itemize}

\header{Step 3: Trace Link Creation.}
Next, we manually examined each documentation segment against the extracted code artifacts to establish \textbf{trace links}. For every segment, we performed a multi-faceted analysis by:

\begin{itemize}
  \item \textbf{Identifying Explicit Mentions:} Searching for direct references by name, such as ``Crawler constructor'', ``\texttt{max\_concurrent\_requests} setting'', or ``\texttt{\_\_init\_\_} method.''
  \item \textbf{Analyzing Implicit Relationships:} Examining textual descriptions and usage examples to identify code artifacts that are used, configured, or demonstrated even if not explicitly named in the narrative.
  \item \textbf{Characterizing and Justifying:} For each potential link, documenting the nature of the relationship (e.g., ``describes configuration of'', ``provides default value for'', ``example instantiates'') and providing a concise justification based on evidence from the documentation and corresponding code.
  \item \textbf{Identifying Trace Chains:} Recognizing sequences where a documentation segment links to one artifact which, in turn, relates to another (for example, a segment might reference a configuration parameter in a method that falls back to a class constant). These chains represent a pathway of understanding from the documentation to deeper implementation details.
\end{itemize}

\emph{Running Example:} For our example segment, trace links would be created to:
\begin{itemize}
  \item \texttt{Crawler.\_\_init\_\_}: Relationship: ``Describes configuration parameter (\texttt{max\_concurrent\_\\requests}) used by this constructor''. Justification: The documentation explains the parameter, and the example shows it passed via \texttt{crawler\_settings} to the \texttt{Crawler} constructor, which then assigns it internally.
  \item \texttt{Crawler.DEFAULT\_CONCURRENCY}: Relationship: ``Specifies the default value for the concept implemented by this constant''. Justification: The text states ``The default is 5'', directly corresponding to the constant used as a fallback in \texttt{\_\_init\_\_}.
  \item \texttt{Crawler} (Class): Relationship: ``Provides a usage example instantiating this class''. Justification: The code example explicitly creates an instance with \texttt{my\_crawler = Crawler(...)}.
\end{itemize}
A trace chain here is: Segment~A $\rightarrow$ \texttt{Crawler.\_\_init\_\_} $\rightarrow$ \texttt{Crawler.DEFAULT\_CONCURRENCY}.

\header{Step 4: Quality Control \& Verification.}
All proposed links underwent a manual verification pass to confirm they were valid and accurately reflected the relationships described in the text. Inter-rater agreement between the first and second authors was Cohen's $\kappa = 0.94$, indicating almost-perfect agreement~\cite{landis1977measurement}. Each proposed trace link was reviewed against several criteria:
\begin{itemize}
\item \textbf{Granularity Consistency:} Does the link connect a documentation segment to a code artifact at the defined level (Class, Method, Attribute)?
\item \textbf{Justification Clarity \& Accuracy:} Is the written explanation for the link clear, technically correct, and well-supported by the content of the segment and the artifact?
\item \textbf{Trace Validity:} Does the link represent a meaningful semantic connection relevant for software understanding or maintenance (e.g., configuration, usage example, implementation detail description)? Are identified trace chains logical pathways?
\item \textbf{Completeness Check:} Considering the segment, are there any obvious, significant links (especially implicit ones revealed by examples) that were missed?
\end{itemize}
Links deemed invalid were either corrected or removed. This review also ensured that only the publicly relevant portions of code (e.g., classes, methods, class-level attributes, etc.) were linked, thus avoiding noise from private implementation details or purely internal variables.

\emph{Running Example:} In the verification process, we would check if the links created in Step 3 for the concurrency segment are accurate (e.g., does \texttt{\_\_init\_\_} actually use the parameter? Is 5 truly the default value constant?). We would assess if the justifications are understandable and confirm that the granularity is correct (linking to the method and attribute, not just arbitrary lines).

\header{Step 5: Dataset Structure Finalization.}
Finally, the verified data was organized into a structured format suitable for experimental use. Each record in the dataset corresponds to one documentation segment. It contains the segment's text, its metadata (source file, location), and a list of all verified code artifacts traced to it. For each linked artifact, the record stores detailed information derived from the previous steps: the artifact's identifier (e.g., \texttt{src/crawler.py::Crawler.\_\_init\_\_}), its type, location, source code, the relationship description, the granularity level of the trace, and any relevant trace chain information. This structured format facilitates parsing the data for experiments and enables various types of analysis on the nature and distribution of trace links.

\emph{Running Example:} The final record for our example segment would bundle the Markdown text, its metadata, and an array containing entries for the links to \texttt{Crawler.\_\_init\_\_}, \texttt{Crawler.\\DEFAULT\_CONCURRENCY}, and \texttt{Crawler}, each with their respective justifications and other details gathered in Steps 2-4. Figure~\ref{fig:dataset-structure-example} illustrates the structure of a dataset entry, using our running example.

% --- Dataset Structure Example Figure ---

\begin{figure}[htbp] % Keep placement flexibility
  \centering
  \begin{snippetbox}
  % Using MINIMAL lstlisting options to ensure visibility
    % In the figure environment for the JSON:
\begin{lstlisting}[basicstyle=\small\ttfamily, columns=flexible,
    breaklines=true, % NEED this for wrapping
    belowskip=\smallskipamount % ADD this tiny skip below the listing
    % Keep minimal other options for now
]
{
  "document": {
    "text": "### Setting Concurrency Level\n\nTo control the number of simultaneous...\nExample Usage:\n```python\ncrawler_settings = {\n    \"start_url\": \"https://example.com\",\n    \"max_concurrent_requests\": 10\n}\nmy_crawler = Crawler(config=crawler_settings)\n```",
    "location": "docs/config.md",
    "type": "Configuration Section"
  },
  "artifacts": [
    {
      "title": "Crawler.__init__",
      "location": "src/crawler.py",
      "content": "def __init__(self, config):\n    self.settings = config\n    self.concurrency = self.settings.get(...)",
      "type": "Method",
      "relationship": "Describes configuration parameter (max_concurrent_requests) used by this constructor.",
      "traceability_granularity": "Method-level",
      "trace_chain": "docs/config.md -> Crawler"
    },
    {
      "title": "Crawler.DEFAULT_CONCURRENCY",
      "location": "src/crawler.py",
      "content": "DEFAULT_CONCURRENCY = 5",
      "type": "Attribute",
      "relationship": "Specifies the default value for the concept implemented by this constant.",
      "traceability_granularity": "Statement-level",
      "trace_chain": "docs/config.md -> Crawler.__init__ -> Crawler.DEFAULT_CONCURRENCY"
    },
    {
      "title": "Crawler",
      "location": "src/crawler.py",
      "content": "class Crawler:\n    DEFAULT_CONCURRENCY = 5 ...",
      "type": "Class",
      "relationship": "Provides usage example instantiating this class.",
      "traceability_granularity": "Class-level",
      "trace_chain": "docs/config.md -> Crawler"
    }
  ]
}
\end{lstlisting} % belowskip applies here
  \end{snippetbox}
  \caption{Illustrative structure of a dataset entry using the running example.}
  \label{fig:dataset-structure-example}
\end{figure}

\vspace{\baselineskip}
We repeated this workflow for all documentation segments across both repositories, yielding two distinct datasets. Table~\ref{tab:dataset-stats} summarizes the key characteristics of each dataset. In \texttt{Crawl4AI}, we obtained 112 documentation segments and 29 code artifacts, resulting in a total of 645 trace links. In contrast, \texttt{Unity Catalog} provided a more focused dataset with 32 segments and 76 code artifacts, yielding 155 trace links. On average, the \texttt{Crawl4AI} dataset produced about 5.8 traces per segment compared to 4.8 in \texttt{Unity Catalog}. In terms of granularity, \texttt{Crawl4AI} exhibits a diverse distribution with 464 class-level, 98 method-level, and 83 statement-level links. For \texttt{Unity Catalog}, however, only 13 class-level and 142 method-level trace links were identified; no statement-level links were found, which is attributable to the nature of its API documentation that predominantly references higher-level code constructs. This variation in scale and density across the two datasets provides a rich basis for evaluating LLM traceability capabilities.

\begin{table}
\caption{Dataset Characteristics}
\label{tab:dataset-stats}
\begin{tabular}{lrr}
\hline
\textbf{Characteristic} & \textbf{Crawl4AI} & \textbf{Unity Catalog} \\
\hline
Document Segments & 112 & 32 \\
Code Artifacts & 29 & 76 \\
\textbf{Total Traces} & \textbf{645} & \textbf{155} \\
\hline 
\textbf{Granularity Distribution} & & \\
\quad Class-level & 464 & 13 \\
\quad Method-level & 98 & 142 \\
\quad Statement-level & 83 & 0 \\
\hline
\end{tabular}
\end{table}

\begin{figure}[htbp]
\centering
\begin{snippetbox}
\begin{codeverbatim}
cached_content = {
    "available_artifacts": [
        {
            "artifact_id": artifact["artifact_id"],
            "title": artifact["title"],
            "location": artifact["location"],
            "content": artifact["content"]
        }
        for artifact in request["artifact_list"]
    ],
    "directory_tree": request["directory_tree"]
}

prompt_dictionary = {
    "task": "Documentation to Code Traceability Analysis",
    "description": "Analyze documentation to identify relevant 
    code artifacts and relationships",
    "document": {
        "text": "<documentation_text>",
        "location": "<file_location>"
    },
    "instructions": {
        "steps": [
            "1. Analyze specific text for traces",
            "2. Identify explicit/implicit mentions",
            "3. Explain relationships",
            "4. Construct trace chains"
        ]
    }
}
\end{codeverbatim}
\end{snippetbox}
\caption{Simplified prompt structure for documentation-to-code traceability.}
\label{fig:prompt-structure}
\end{figure}

\subsection{Experimental Setup}
\label{sec:exp_setup}

\header{Overview.}
In this section, we describe the experimental methodology used to evaluate the traceability capabilities of the selected LLMs. We outline the dataset preparation, explain the evaluation strategy (including data ordering and prompting approach), define the baselines chosen for comparison, and clarify the metrics and additional analyses performed.

\header{Datasets.}
We conducted our evaluation using the two datasets described in Section~\ref{sec:project_selection}: \emph{Crawl4AI} (112 documentation segments, 29 code artifacts) and \emph{Unity Catalog} (32 documentation segments, 76 code artifacts). Each documentation segment is assessed individually against all available code artifacts (one-to-many strategy, other strategies are discussed in Section \ref{subsec:context-analysis}).

\header{Evaluation Procedure.}
To provide more robust performance estimates, we repeat each experiment five times. Each experiment involves:
\begin{itemize} 
    \item Shuffling the order of documentation segments. 
    \item Shuffling the order of code artifacts presented to the LLM. 
    \item Feeding each segment (with its list of artifacts) to the LLM and collecting its predicted links. 
    \item Aggregating all predictions across segments to compute precision, recall, and F1 for that run. 
\end{itemize}
These steps help to account for potential sensitivity to input order and allow us to report average performance across randomized evaluation runs.

\header{Prompt Structure.}
% We created a structured prompt for the documentation-to-code traceability task, as depicted in Figure~\ref{fig:prompt-structure}. The prompts explicitly instruct the LLMs to identify both explicit and implicit references within the given documentation segment, restricting trace analysis solely to provided artifacts. Each prompt contains two main components:
We created a structured prompt for the documentation-to-code traceability task, as depicted in Figure~\ref{fig:prompt-structure}. Unlike traditional single-text prompts, our format uses a structured input schema to provide the LLM with separate, well-organized fields for instructions, context (available code artifacts), and the documentation segment under analysis \cite{anthropicpromptingoverview, openaipromptingoverview}. This design ensures clarity, repeatability, and control over the LLM’s attention. The prompts explicitly instruct the LLMs to identify both explicit and implicit references within the given documentation segment, restricting trace analysis solely to the provided artifacts. Each prompt contains two main components

\begin{itemize}
    \item \texttt{cached\_content}: A JSON object listing available artifacts (their titles, content, and locations) and the directory tree structure of the repository, providing relevant context.
    \item \texttt{prompt\_dictionary}: Detailed instructions directing the LLMs' analysis, including how to identify relationships, differentiate between explicit and implicit references, and correctly format trace chains (e.g., \texttt{docs.md $\rightarrow$ ClassA $\rightarrow$ MethodB}).
\end{itemize}

It is important to note that Figure~\ref{fig:prompt-structure} shows a simplified view for clarity and brevity. The actual prompt employed is considerably more elaborate, featuring numerous detailed steps (e.g., specifying granularity rules, defining relationship types, formatting trace chains precisely) and strict output schema enforcement, as necessitated by the complexity of the traceability task. Complete versions of the prompts are provided in the Appendix.

\header{LLM Settings and API Details.}
We evaluated three LLMs: \emph{Claude 3.5 Sonnet}, \emph{GPT-4o}, and \emph{o3-mini}, as described in detail in Section~\ref{sec:llm_selection}. Default parameters were employed in each case, except that the parameter \texttt{reasoning\_effort} was explicitly set to \texttt{high} for \emph{o3-mini} to maximize its reasoning performance. API requests were made directly through the official providers (Anthropic and OpenAI). For efficiency, we leveraged built-in request caching: OpenAI enables caching by default for all API calls, while for Anthropic, it was explicitly enabled for our experiments.

\header{Baseline Methods.}
We compared the LLM results against three widely used baseline methods in software traceability:

\begin{itemize}
    \item \textbf{TF-IDF}: computes cosine similarity between TF-IDF vectorized documentation segment and code artifact \cite{antoniol2002recovering}.
    \item \textbf{BM25}: employs a probabilistic token-matching approach (Okapi BM25) to rank code artifacts given a documentation segment \cite{robertson2009probabilistic}.
    \item \textbf{CodeBERT}: utilizes pretrained CodeBERT embeddings trained specifically on software texts and code. Due to its 512-token limit, CodeBERT applies a sliding-window approach for longer inputs, averaging embeddings across windows before computing cosine similarities \cite{feng2020codebert}.
\end{itemize}

% We performed threshold tuning for each baseline individually, conducting a grid search in 0.05 increments to select the optimal similarity threshold based on F1-score.
Each baseline method outputs a similarity score between a documentation segment and a code artifact—typically in the range [0,1] for TF-IDF and CodeBERT (cosine similarity), and a non-normalized score for BM25. To convert these continuous scores into binary trace links, we perform threshold tuning for each method individually. Specifically, we conduct a grid search in 0.05 increments and select the threshold that maximizes F1-score on the evaluation set. This represents an optimistic estimate of baseline performance, since the thresholds are selected post hoc based on the same data being evaluated.

\header{Metrics and Granularity Analysis (RQ1).}
We measure standard Information Retrieval metrics: precision, recall, and F1-score \cite{antoniol2002recovering, marcus2003recovering}. Additionally, we report metrics grouped by artifact granularity (class-level, method-level, and statement-level), providing deeper insight into how performance varies with artifact detail \cite{blasco2020fine, lucia2007recovering}. These granular metrics help clarify the strengths and weaknesses of each approach at different traceability levels.

\header{Relationship Explanation (RQ2).}
For RQ2, we evaluated the quality of relationship explanations provided by the LLMs. Given the substantial volume of generated links, we used an LLM-based evaluation approach (LLM as a judge) \cite{wang2025can, ahmed2024can}, classifying explanations into three accuracy categories: \emph{correct}, \emph{partially correct}, or \emph{incorrect}. To validate the reliability of this automated evaluation, a representative subset was manually reviewed and verified. More details on this are found in Section~\ref{sec:rq2}.

\header{Trace Chain Analysis (RQ3).}
For RQ3, we analyzed trace chains generated by LLMs, comparing predicted chains against ground truth. We classified predicted chains based on how closely they matched actual trace pathways, using defined correctness categories: \textit{complete match}, \textit{partial match} (with subcategories), or \textit{incorrect}. Section~\ref{sec:rq3} describes these categories and analysis details further.

\subsection{Reproducibility and Ethical Considerations}
\label{sec:reproducibility_ethics}

To ensure the reproducibility of our study, all experiments were implemented in Python~3.11. The code, datasets, analysis scripts, and prompt templates used are publicly available in our replication package\footnote{\url{https://github.com/Alor-e/evaluating-llm-doc-code-traceability}}. This package includes detailed instructions for reproducing our results, and we have documented our experimental procedures following standard research practices.

Throughout the study, we adhered to ethical guidelines. This included respecting the licenses of the software projects used in our datasets, complying with the terms of service and usage policies provided by the LLM vendors, and ensuring that no sensitive or personal data was included or exposed during dataset creation or experimentation.

\section{RQ1: Trace Link Accuracy}
\label{sec:rq1}
For RQ1, \textbf{\emph{\rqi}}, we aim to examine if LLMs can establish traceability effectively between document and code elements. We present our motivation, approach, and results.

\subsection{Motivation}
\label{sec:rq1_motivation}
While LLMs demonstrate strong capabilities in understanding both natural language and code \cite{chen2021evaluating, karanikolas2023large}, their ability to create accurate trace links between software documentation and source code has not been systematically studied. Establishing these links is essential for core software engineering tasks like change impact analysis, program comprehension, and maintaining documentation consistency with evolving code \cite{cleland2014software, uddin2015api}. Traditional automated methods, however, often struggle to bridge the semantic gap between high-level documentation and specific code constructs needed for these tasks, leading to inaccurate or incomplete results \cite{antoniol2002recovering, lucia2007recovering}. LLMs offer a promising alternative due to their advanced semantic understanding \cite{fan2023large}. Therefore, assessing their accuracy is essential to determine if LLMs are useful for improving documentation to code traceability.

\subsection{Approach}
\label{sec:rq1_approach}

To answer this RQ, we evaluate the traceability capabilities of the selected LLMs: Claude Sonnet 3.5, GPT-4o, and o3-mini. Each LLM is asked to analyze documentation segments to identify relevant code artifacts. This analysis considered both explicitly named artifacts mentioned in the text and implicitly related artifacts inferred from contextual clues, such as code usage examples, presented within the documentation.

For each documentation segment, we provided the LLMs with relevant contextual information: the document text itself, its location within the file. For the code artifacts, we provided the LLMs with their location within the directory, the code itself, and the project's directory structure. This rich context enabled the LLMs to understand both local and project-wide relationships.

Our evaluation process incorporated multiple measures to ensure reliable results. We conducted five runs with different random seeds and input orderings to account for potential variations in LLM responses. To compare fairly, we gave both the LLMs and the baseline methods the same inputs: the document segments and the list of code artifacts to check against using a one-to-many matching strategy. We assessed the overall accuracy of trace link identification using standard Information Retrieval metrics: Precision, Recall, and F1-score. 

To further understand the LLMs performance, we employed a mixed-methods approach. For all incorrect trace links (False Positives), we performed a qualitative analysis through manual review, categorizing errors into patterns such as context boundary violations (i.e., linking a documentation segment to code mentioned elsewhere in the file but outside the segment's immediate context; see Section \ref{sec:rq1_results}) and unsupported implicit assumptions (i.e., inferring links without sufficient evidence, see Section \ref{sec:rq1_results}). For missed trace links (False Negatives), we performed a quantitative analysis comparing their characteristics against those of correctly identified trace links to identify factors associated with successful or failed trace recovery, examining attributes including document length, presence of code examples, and trace link granularity.

\subsection{Results}
\label{sec:rq1_results}

\begin{table}
\caption{Documentation-to-Code Traceability Performance Across LLMs, Baseline Models and Datasets}
\label{tab:traceability_results}
\centering
\begin{tabular}{@{}ll
                S[table-format=3.1] % Precision
                S[table-format=3.1] % Recall
                S[table-format=3.1] % F1
               @{}}
\toprule
\textbf{Dataset} & \textbf{Model} & {\textbf{Precision (\%)}} & {\textbf{Recall (\%)}} & {\textbf{F1 (\%)}} \\
\midrule
\multirow{16}{*}{\textbf{Crawl4AI}}
 & \textbf{Claude Sonnet 3.5} & 95.3 & 68.0 & \textbf{79.4} \\
 & \quad Class     & 93.7 & 57.3 & 71.1 \\
 & \quad Method    & 97.8 & 100.0 & 98.9 \\
 & \quad Statement & 98.0 & 90.5 & 94.0 \\
\cmidrule{2-5}
 & \textbf{GPT-4o} & 96.8 & 53.0 & 68.5 \\
 & \quad Class     & 96.4 & 41.7 & 58.2 \\
 & \quad Method    & 96.4 & 98.6 & 97.5 \\
 & \quad Statement & 99.2 & 62.2 & 76.5 \\
\cmidrule{2-5}
 & \textbf{o3-mini} & \textbf{98.5} & 58.0 & 73.0 \\
 & \quad Class     & 98.4 & 45.7 & 62.4 \\
 & \quad Method    & 97.6 & 98.4 & 98.0 \\
 & \quad Statement & 100.0 & 79.3 & 88.4 \\
\cmidrule{2-5}
 & \textbf{TF-IDF}  & 50.3 & 58.7 & 54.2 \\
 & \textbf{BM25}    & 33.4 & 65.0 & 44.1 \\
 & \textbf{CodeBERT}& 22.8 & \textbf{87.7} & 36.2 \\
\midrule
\multirow{13}{*}{\textbf{Unity Catalog}}
 & \textbf{Claude Sonnet 3.5} & 87.0 & 74.8 & \textbf{80.4} \\
 & \quad Class     & 97.3 & 100.0 & 98.6 \\
 & \quad Method    & 85.8 & 72.5 & 78.6 \\
\cmidrule{2-5}
 & \textbf{GPT-4o} & 99.7 & 53.2 & 69.3 \\
 & \quad Class     & 100.0 & 18.5 & 30.1 \\
 & \quad Method    & 99.7 & 56.3 & 72.0 \\
\cmidrule{2-5}
 & \textbf{o3-mini} & \textbf{100.0} & 47.3 & 64.2 \\
 & \quad Class     & 60.0 & 6.2 & 11.0 \\
 & \quad Method    & 100.0 & 51.1 & 67.6 \\
\cmidrule{2-5}
 & \textbf{TF-IDF}  & 77.6 & 62.6 & 69.3 \\
 & \textbf{BM25}    & 12.9 & 40.0 & 19.5 \\
 & \textbf{CodeBERT}& 6.4  & \textbf{100.0} & 12.0 \\
\bottomrule
\multicolumn{5}{@{}l@{}}{\small Note: Bold values indicate best overall performance for each metric within each dataset.}\\
\end{tabular}
\end{table}

Table~\ref{tab:traceability_results} presents performance metrics across the selected LLMs, baseline models, and datasets, demonstrating both the strengths and limitations of different approaches. Within each dataset, the best-performing model (confirmed by the \emph{Scott-Knott ESD} test \cite{tantithamthavorn2017mvt, tantithamthavorn2018optimization}) for each metric (Precision, Recall, F1) is highlighted in bold.  For brevity, we omit the baselines' granularity breakdown here, but those detailed results can be found in the Appendix. 

\header{Traceability Performance.}
The evaluated LLMs demonstrate overall F1 scores (across all granularities) ranging from 64.2\% to 80.4\% across both datasets, indicating a moderate to strong performance in identifying traceability links. A key strength of the LLMs lies in their overall precision, which is consistently high (above 87\%) and substantially outperforms the baseline methods. However, while overall precision is uniformly high, their overall recall varies significantly (ranging from 47\% to 75\% across the models and datasets). This divergence in recall is the primary factor driving the differences in their overall F1 scores.

We observe consistent performance patterns among the LLMs across the datasets. Claude Sonnet 3.5 achieves the best F1 score, demonstrating the most effective balance between high overall precision and the highest overall recall among the evaluated LLMs. Conversely, o3-mini achieves the highest overall precision, which could be attributed to its reasoning capabilities. The baseline CodeBERT achieves the highest overall recall, but this is accompanied by very low precision, indicating that it identifies almost all possible links, including a large number of incorrect ones. It is also noteworthy that on the Unity Catalog dataset, the TF-IDF baseline achieves an F1 score competitive with GPT-4o and superior to o3-mini on this specific dataset.

Furthermore, performance varies across the granularity of the code artifacts being traced, especially regarding recall compared to precision. Method-level tracing generally achieves higher F1 scores across the models and datasets compared to class-level tracing (e.g., 97.5\%--98.9\% vs. 58.2\%--71.1\% in Crawl4AI), with statement-level performance typically falling in between (e.g., 76.5\%--94.0\% in Crawl4AI); however, exceptions exist, such as Claude Sonnet 3.5's strong class-level performance in Unity Catalog.

\header{Analyzing Incorrect Links (False Positives).}
\label{sec:rq1_fp_analysis}

\begin{table}
\caption{False Positive Error Pattern Analysis Across LLMs and Datasets}
\label{tab:error_patterns}
\centering
\sisetup{round-mode=places, round-precision=1, table-format=3.1} % Setup siunitx formatting (adjusted table-format)
\begin{tabular}{@{}l l l S@{}} % Dataset, Model, Error Pattern, Percentage
\toprule
\textbf{Dataset} & \textbf{Model} & \textbf{Error Pattern} & {\textbf{Frequency (\%)}} \\
\midrule
\multirow{16}{*}{\textbf{Crawl4AI}} % Span 15 rows: 3 models * (4 error types + 1 total)
 & \multirow{5}{*}{Claude Sonnet 3.5} % Span 5 rows for this model
 & Implicit Assumption (IAE) & 61.3 \\
 & & Phantom Link (PAL) & 30.7 \\
 & & Architecture Pattern Bias (APB) & 0.0 \\
 & & Implementation Overlink (IOL) & 8.0 \\
 \cmidrule(l){3-4} % Rule under error patterns for this model
 & & \textit{Total False Positives} & {\textit{150}} \\ % Absolute count for context
 \cmidrule(l){2-4} % Rule separating models
 & \multirow{5}{*}{GPT-4o} % Span 5 rows for this model
 & Implicit Assumption (IAE) & 98.2 \\ % 55/56
 & & Phantom Link (PAL) & 1.8 \\  % 1/56
 & & Architecture Pattern Bias (APB) & 0.0 \\
 & & Implementation Overlink (IOL) & 0.0 \\
 \cmidrule(l){3-4} % Rule under error patterns for this model
 & & \textit{Total False Positives} & {\textit{56}} \\  % Absolute count for context
 \cmidrule(l){2-4} % Rule separating models
 & \multirow{5}{*}{o3-mini} % Span 5 rows for this model
 & Implicit Assumption (IAE) & 52.6 \\ % 20/38
 & & Phantom Link (PAL) & 0.0 \\
 & & Architecture Pattern Bias (APB) & 0.0 \\
 & & Implementation Overlink (IOL) & 47.4 \\ % 18/38
 \cmidrule(l){3-4} % Rule under error patterns for this model
 & & \textit{Total False Positives} & {\textit{38}} \\  % Absolute count for context
\midrule % Separator between datasets
\multirow{17}{*}{\textbf{Unity Catalog}} % Span 15 rows
 & \multirow{5}{*}{Claude Sonnet 3.5} % Span 5 rows for this model
 & Implicit Assumption (IAE) & 51.9 \\ % 67/129
 & & Phantom Link (PAL) & 15.5 \\ % 20/129
 & & Architecture Pattern Bias (APB) & 32.6 \\ % 42/129
 & & Implementation Overlink (IOL) & 0.0 \\
 \cmidrule(l){3-4} % Rule under error patterns for this model
 & & \textit{Total False Positives} & {\textit{129}} \\ % Absolute count for context
 \cmidrule(l){2-4} % Rule separating models
 & \multirow{5}{*}{GPT-4o} % Span 5 rows for this model
 & Implicit Assumption (IAE) & 100.0 \\ % 1/1 (Assuming the single error was IAE)
 & & Phantom Link (PAL) & 0.0 \\
 & & Architecture Pattern Bias (APB) & 0.0 \\
 & & Implementation Overlink (IOL) & 0.0 \\
 \cmidrule(l){3-4} % Rule under error patterns for this model
 & & \textit{Total False Positives} & {\textit{1}} \\ % Absolute count for context
 \cmidrule(l){2-4} % Rule separating models
 & \multirow{5}{*}{o3-mini} % Span 5 rows for this model
 & Implicit Assumption (IAE) & 0.0 \\ % 0/0 - Use 0.0 for consistency with siunitx format
 & & Phantom Link (PAL) & 0.0 \\
 & & Architecture Pattern Bias (APB) & 0.0 \\
 & & Implementation Overlink (IOL) & 0.0 \\
 \cmidrule(l){3-4} % Rule under error patterns for this model
 & & \textit{Total False Positives} & {\textit{0}} \\ % Absolute count for context
\bottomrule
\multicolumn{4}{@{}p{0.9\linewidth}@{}}{\footnotesize Note: The frequency shows the proportion of each error type relative to the total false positives for that specific model/dataset combination.} \\
\end{tabular}
\end{table}

Table~\ref{tab:error_patterns} presents the frequency of error patterns across LLMs and datasets. Note that GPT-4o and o3-mini showed very high precision scores and thus minimal (0 or 1) total false positives. Below, we discuss each error category, arranged from most to least frequent, highlighting specific reasons behind each type of failure.

\textit{Implicit Assumption Errors (IAE)} occurred when LLMs inferred relationships without sufficient evidence, often due to reliance on naming conventions or their pretrained knowledge of common software patterns. IAEs were the most frequent errors across all LLMs. For example, Claude Sonnet 3.5 mistakenly assumed a class inheritance between \texttt{CosineStrategy} and \texttt{ChunkingStrategy} classes in the Crawl4AI dataset based purely on naming similarity, even though no explicit inheritance existed in the code.

\textit{Phantom Link (PAL)} errors were generated when LLMs traced documentation to code artifacts referenced implicitly or explicitly in the documentation but not present within the provided code artifact set. Claude Sonnet 3.5 frequently produced PAL errors, such as incorrectly linking documentation to the method \texttt{NlpSentenceChunking.chunk()}, which was mentioned in documentation examples in the Crawl4AI dataset but absent from the provided artifacts. In contrast, GPT-4o and o3-mini rarely encountered this issue.

\textit{Architecture Pattern Bias (APB)} errors arose from LLMs' tendency to overgeneralize architectural patterns observed in specific parts of the codebase, applying them incorrectly to unrelated areas. Such errors were only observed in Claude Sonnet 3.5 for the Unity Catalog dataset, where the model incorrectly assumed that all services adhered to a specific architectural pattern (e.g., \texttt{ModelService} $\rightarrow$ \texttt{ModelRepository}). Services employing different architectures (e.g., temporary credential services in the dataset) often triggered APB errors.

\textit{Implementation Overlink (IOL)} errors resulted when LLMs inappropriately traced documentation to internal implementation details, violating abstraction boundaries. Claude Sonnet 3.5 and o3-mini generated these errors in the Crawl4AI dataset, incorrectly linking documentation segments to private helper methods or internal functions not intended for external documentation reference. GPT-4o notably avoided such errors entirely.

\header{Analyzing Missed Links (False Negatives).}
\label{sec:rq1_fn_analysis}

\begin{table}
\caption{Comparison of Document Characteristics Impact on Trace Recovery Across LLMs and Datasets}
\label{tab:trace_characteristics}
\centering
\small
% Updated column specification: removed last two 'l'
\begin{tabular}{@{}lll S[table-format=3.1] S[table-format=1.1]@{}}
\toprule
\textbf{Dataset} & \textbf{Model} & \textbf{Characteristic} & {\textbf{TP}} & {\textbf{FN}} \\ % Removed Magnitude/p-value headers
\midrule

\multirow{18}{*}{\textbf{Crawl4AI}}

 & \multirow{6}{*}{Claude Sonnet 3.5}
 & Explicit Mentions (\%) & 44.6 & 9.1 \\
 & & Usage Examples (\%) & 94.7 & 97.9 \\
 & & Avg. Document Length (words) & 85.3 & 79.0 \\ 
 & & Class-level (\%) & 60.9 & 96.2 \\
 & & Method-level (\%) & 22.3 & 0.0 \\
 & & Statement-level (\%) & 16.9 & 3.8 \\
\cmidrule{2-5}

 & \multirow{6}{*}{GPT-4o}
 & Explicit Mentions (\%) & 44.8 & 18.3 \\ 
 & & Usage Examples (\%) & 94.3 & 97.9 \\ 
 & & Avg. Document Length (words) & 83.4 & 82.0 \\
 & & Class-level (\%) & 56.7 & 89.4 \\
 & & Method-level (\%) & 28.5 & 0.5 \\
 & & Statement-level (\%) & 14.8 & 10.1 \\
\cmidrule{2-5}

 & \multirow{6}{*}{o3-mini}
 & Explicit Mentions (\%) & 49.2 & 9.3 \\
 & & Usage Examples (\%) & 93.8 & 98.9 \\
 & & Avg. Document Length (words) & 86.1 & 79.0 \\
 & & Class-level (\%) & 56.7 & 93.1 \\
 & & Method-level (\%) & 25.9 & 0.6 \\ 
 & & Statement-level (\%) & 17.5 & 6.3 \\
\midrule

\multirow{16}{*}{\textbf{Unity Catalog}}

 & \multirow{5}{*}{Claude Sonnet 3.5}
 & Explicit Mentions (\%) & 87.9 & 97.4 \\ 
 & & Parameters/Returns (\%) & 60.4 & 23.7 \\ 
 & & Avg. Document Length (words) & 83.2 & 81.5 \\ 
 & & Class-level (\%) & 11.2 & 0.0 \\  
 & & Method-level (\%) & 88.8 & 100.0 \\
\cmidrule{2-5}

 & \multirow{5}{*}{GPT-4o}
 & Explicit Mentions (\%) & 96.1 & 83.7 \\ 
 & & Parameters/Returns (\%) & 68.7 & 37.7 \\ 
 & & Avg. Document Length (words) & 89.0 & 75.7 \\ 
 & & Class-level (\%) & 2.9 & 14.6 \\ 
 & & Method-level (\%) & 97.1 & 85.4 \\
\cmidrule{2-5} 

 & \multirow{5}{*}{o3-mini}
 & Explicit Mentions (\%) & 97.5 & 83.7 \\ 
 & & Parameters/Returns (\%) & 66.4 & 43.2 \\ 
 & & Avg. Document Length (words) & 88.8 & 78.0 \\ 
 & & Class-level (\%) & 1.1 & 15.1 \\ 
 & & Method-level (\%) & 98.9 & 84.9 \\
\bottomrule
\end{tabular}
\end{table}

To understand why LLMs sometimes fail to identify correct trace links (False Negatives), we compared the characteristics of the documentation segments between correctly identified links (True Positives) and missed links (False Negatives). Table~\ref{tab:trace_characteristics} presents this comparison across several key characteristics of each dataset. We specifically examined the presence of \textit{explicit mentions} (whether the documentation directly names the code artifact), \textit{usage examples} (whether there are segments containing code snippets within fenced blocks). We also considered \textit{structured sections} such as \texttt{Parameters} and \texttt{Returns} specific to the Unity Catalog dataset (as commonly found in API-style documentation). Finally, we considered the different levels of granularity of the trace links (i.e., Class-level, Method-level, and Statement-level, where applicable).
The results differ notably between the two datasets due to their distinct documentation styles.

\textit{Crawl4AI.} This dataset uses narrative-style documentation, often featuring illustrative usage examples. The results show that explicit mentions of code artifacts greatly influence trace link success: segments with explicit mentions were consistently more successful compared to missed links across the LLMs (45--49\% of TPs vs only 9--18\% of FNs). Importantly, missed links predominantly occur at the class level (at least 89\% of all FNs), highlighting a clear difficulty for the LLMs in identifying class-level information in less structured, narrative-style documentation. Method-level and statement-level links rarely resulted in misses, aligning with the previously observed higher recall at these granularities (as seen in Table~\ref{tab:traceability_results}).
% Usage examples are very common in both successful and missed segments (over 94\%) and does not seem to be a distinguishing characteristic. Similary, Avg. Document Length (words).

\textit{Unity Catalog.} Unlike Crawl4AI, Unity Catalog features structured API reference documentation without explicit usage examples or statement-level traces. The critical factor influencing traceability is the existence of structured sections (\texttt{Parameters} and \texttt{Returns}) (60--69\% of TPs vs 24--43\% of FNs). Document segments containing these structured sections are more likely to be linked correctly, while those lacking such structure see considerably higher miss rates. 
% Explicit mentions are common in both successfully identified links and missed ones (ranging from 84--98\% across TPs and FNs) and does not seem to be a distinguishing characteristic.
% Missed trace links frequently occur at the method level, reflecting the difficulty models face when distinguishing among multiple similar API methods in the absence of clearly structured contextual information.

\textit{Cross-dataset Insights.} The differing patterns in these two datasets suggest that successful traceability heavily relies on the dominant type of contextual cue provided by the documentation. In Crawl4AI (narrative-style documentation), explicit mentions of artifacts enhance recall (due to lower false negatives), especially at higher abstraction levels (i.e., classes). In Unity Catalog (structured API documentation), where explicit mentions are nearly always present, additional structural cues such as parameter and return tables is decisive in correctly establishing trace links.

\begin{tcolorbox}[colback=white,colframe=gray!50, coltitle=black, arc=3pt,title=\textbf{Key Findings for RQ1}]
\noindent
\textbf{Traceability Performance.} 

All evaluated LLMs achieved high precision (87\%--100\%) with F1‑scores between 64.2\% and 80.4\%, confirming their ability to create trace links; \emph{Claude~Sonnet~3.5} obtained the best F1 on both datasets.

\medskip
\textbf{Common Error Patterns.} False positives stem mainly from  

(1)~naming‑based assumptions (\textbf{IAE}),
(2)~phantom links to non‑existent artifacts (\textbf{PAL}),
(3)~over‑generalised architectural patterns (\textbf{APB}), and
(4)~implementation overlinking to internal details (\textbf{IOL}).

\medskip
\textbf{Documentation Cues Drive Recall.}  
In Crawl4AI (narrative-style documentation), \emph{explicit mentions} of a code artifact are the strongest predictor of recall; in Unity Catalog (structured API documentation), the presence of structural cues, \texttt{Parameters} /\ \texttt{Returns} tables, plays the same role.
\end{tcolorbox}

\section{RQ2: Trace Link Explanation}
\label{sec:rq2}
For RQ2, \textbf{\emph{\rqii}} we aim to analyze if LLMs can understand relationships between document and code elements. We present our motivation, approach, and results.

\subsection{Motivation}
\label{sec:rq2_motivation}
In RQ1, we found that LLMs can accurately identify trace links between documentation segments and their corresponding code elements. While identifying trace links is fundamental, understanding the specific nature of these relationships is crucial for software maintenance and comprehension. Simple binary links (i.e., traceable vs non-traceable) do not capture how code implements, extends, or relates to documented functionality. We need to evaluate whether LLMs can provide meaningful explanations of these relationships that match human understanding.

\subsection{Approach}
\label{sec:rq2_approach}

To evaluate the selected LLMs' ability to explain relationships, we designed a two-part assessment process. First, we collected relationship explanations from each LLMs for all correctly identified trace links from RQ1. Then, due to the large volume of generated links, we developed an evaluation framework using a separate LLM as a judge \cite{wang2025can, ahmed2024can} to assess these explanations against our ground truth. The selected model for the LLM as a judge is Claude Sonnet 3.5 (claude-3-5-sonnet-20241022). The prompt for the evaluation can be found in the Appendix.

For each relationship evaluation, we provide the judge LLM with:
\begin{itemize}
    \item The original documentation segment,
    \item The actual code artifact,
    \item The LLMs' generated relationship explanation,
    \item The ground truth relationship description.
\end{itemize}

We asked the judge LLM to classify the generated relationship explanation as correct (accurate and complete relationship description), partially correct (some aspects captured but others missed), and incorrect (failed to capture the true relationship). We deliberately focused only on True Positive cases from RQ1 for two key reasons. First, to evaluate an LLM's ability to explain relationships in a realistic scenario, it must first identify that a relationship exists. Second, using all possible document-code pairs returned by the LLM (i.e., including false positives) would involve asking the LLM to explain relationships where none exist as it is a false positive. In addition to the classification labels (correct, partially correct, and incorrect), we also asked the judge LLM to produce a justification for each classification and error types identified for partially correct and incorrect cases. We then manually aggregated the error types in order to perform our error analysis. We report two complementary metrics: (1) \emph{strict accuracy}, the percentage of explanations the judge classifies as \emph{correct}, and (2) \emph{relaxed accuracy}, the percentage classified as either \emph{correct} or \emph{partially correct}. Relaxed accuracy, therefore, shows how often the LLM captures the gist of the relationship even if some descriptive details are missing.

\subsection{Results}
\label{sec:rq2_results}

\begin{table}
\caption{Relationship Explanation Performance Across LLMs}
\label{tab:relationship_accuracy}
\centering
\begin{tabular}{@{}lrrrr|rr@{}}
\toprule
\textbf{LLM/Dataset} & \textbf{Total} & \textbf{Correct} & \textbf{Partial} & \textbf{Incorrect} & \textbf{Strict Acc.} & \textbf{Relaxed Acc.} \\
\midrule
\textbf{Claude Sonnet 3.5} & & & & & & \\   % header row now spans all columns
Crawl4AI       & 2,200 & 983 & 1,196 & 21 & 44.7\% & 99.0\% \\
Unity Catalog  & 580   & 334 & 245   & 1  & 57.6\% & 99.8\% \\
\midrule
\textbf{GPT-4o} & & & & & & \\                % keeps the rule continuous
Crawl4AI       & 1,687 & 723 & 927   & 37 & 42.9\% & 97.8\% \\
Unity Catalog  & 412   & 200 & 211   & 1  & 48.5\% & 99.8\% \\
\midrule
\textbf{o3-mini} & & & & & & \\              % same idea here
Crawl4AI       & 1,862 & 1,021 & 820  & 21 & 54.8\% & 98.9\% \\
Unity Catalog  & 363   & 258 & 104   & 1  & 71.1\% & 99.7\% \\
\bottomrule
\end{tabular}
\end{table}

Table~\ref{tab:relationship_accuracy} presents the results of our relationship explanations analysis across all three LLMs and both datasets. The results presented are the sum of explanations (TPs) across all runs. The results reveal interesting patterns in how different LLMs understand and explain documentation-code relationships.

\header{Relationship Explanation Performance.} All LLMs demonstrate strong capability in identifying the fundamental nature of relationships, with relaxed accuracy exceeding 97\% across all LLM-dataset combinations. However, their ability to provide both complete and precise explanations (strict accuracy) varies with the majority below 58\%. The o3-mini LLM shows stronger performance in strict accuracy compared to the other LLMs, achieving the highest scores of 54.8\% and 71.1\% for Crawl4AI and Unity Catalog, respectively. This could perhaps be attributed to its reasoning capabilities.

A consistent pattern also emerges across all LLMs: explanation performance on Unity Catalog significantly exceeds that on Crawl4AI, with strict accuracy improvements ranging from 5.6 to 16.3 percentage points. This suggests that Unity Catalog's more structured documentation style facilitates more complete and precise relationship explanations.

\begin{table}
\centering
\footnotesize            % Slightly smaller text
\setlength{\tabcolsep}{3pt} % Tighter space between columns
\caption{Distribution of Error Types Across LLMs with Average Frequencies}
\label{tab:error_types}
\begin{tabular}{@{}lrrrrrr>{\centering\arraybackslash}p{18mm}@{}}
\toprule
 & \multicolumn{2}{c}{\textbf{Claude Sonnet 3.5}} 
 & \multicolumn{2}{c}{\textbf{GPT-4o}} 
 & \multicolumn{2}{c}{\textbf{o3-mini}} 
 & \textbf{Avg. Freq. (\%)} \\   % Ensure all on one line
\cmidrule(lr){2-3} \cmidrule(lr){4-5} \cmidrule(lr){6-7}
\textbf{Error Type} 
 & \multicolumn{1}{c}{Crawl4AI}
 & \multicolumn{1}{c}{Unity C.}
 & \multicolumn{1}{c}{Crawl4AI}
 & \multicolumn{1}{c}{Unity C.}
 & \multicolumn{1}{c}{Crawl4AI}
 & \multicolumn{1}{c}{Unity C.}
 & \\
\midrule
Missing Key Details    & 41.3\% & 42.3\% & 46.2\% & 37.7\% & 46.5\% & 21.9\% & 39.3\% \\
Incomplete Explanation & 50.2\% & 14.2\% & 42.3\% & 21.2\% & 39.0\% & 13.3\% & 30.0\% \\
Implementation Detail  & 0.6\%  & 10.2\% & 0.6\%  & 10.4\% & 3.1\%  & 25.7\% & 8.4\%  \\
Misinterpretation      & 1.6\%  & 8.1\%  & 1.4\%  & 9.0\%  & 1.3\%  & 16.2\% & 6.3\%  \\
Unsupported Assumption & 1.6\%  & 15.9\% & 0.8\%  & 9.0\%  & 1.3\%  & 3.8\%  & 5.4\%  \\
Architectural          & 0.2\%  & 6.1\%  & 0.4\%  & 7.1\%  & 0.2\%  & 12.4\% & 4.4\%  \\
Scope Mismatch         & 3.5\%  & 0.4\%  & 5.6\%  & 0.9\%  & 6.1\%  & 1.9\%  & 3.1\%  \\
Other                  & 0.8\%  & 2.8\%  & 2.7\%  & 4.7\%  & 2.5\%  & 4.8\%  & 3.1\%  \\

\bottomrule
\multicolumn{8}{@{}l@{}}{\small Note: Unity C. is an abbreviation for Unity Catalog. Percentages are calculated relative to the total} \\
\multicolumn{8}{@{}l@{}}{\small error counts for each LLM--dataset combination.}\\
\end{tabular}
\end{table}

\header{Error Pattern Analysis.}
Table~\ref{tab:error_types} reports each error type as a percentage of the total non-correct explanations (i.e., partial or incorrect) for each LLM. The two ``completeness'' categories---\emph{Missing Key Details} (average 39.3\%) and \emph{Incomplete Explanation} (average 30.0\%)---dominate. Despite recognizing the correct linkage, the LLMs frequently omit key details (e.g., stating that \texttt{fetchData()} ``retrieves results'' but failing to mention that the results are paginated) or only partially convey how a code artifact implements or extends the documented functionality. \emph{Implementation Detail} errors (average 8.4\%) occur when explanations drift into low-level, internal specifics (e.g., implementation details of helper functions); such errors are more frequent in o3-mini, which could be attributed to its reasoning capabilities. \emph{Misinterpretation} (average 6.3\%) and \emph{Unsupported Assumption} (average 5.4\%) involve factually incorrect claims or unjustified inferences, such as overstating inheritance relationships. \emph{Architectural} errors (4.4\%) show that LLMs sometimes misapply an observed pattern to unrelated parts of the system. \emph{Scope Mismatch} (average 3.1\%) errors arise when explanations are too broad or too narrow, whereas \emph{Other} (average 3.1\%) captures uncommon, idiosyncratic mistakes. 

\begin{tcolorbox}[colback=white,colframe=gray!50, coltitle=black, arc=3pt,title=\textbf{Key Findings for RQ2}]
\noindent
\textbf{Strong Essential Understanding.} LLMs demonstrate high capability in essential relationship identification,
with relaxed accuracy consistently exceeding 97\% across all LLMs and datasets.

\medskip
\textbf{Weaker Complete Understanding.} In most cases, the LLMs have a strict of below 58\% indicating a much weaker complete understanding.

\medskip
\textbf{Error Characteristics.} Completeness-related errors dominate across all LLMs (average 69.3\% of errors),
with "Missing Key Details" and "Incomplete Explanation" being the primary issues rather than misinterpretation
or incorrect understanding.
\end{tcolorbox}

\section{RQ3: Intermediate Element Detection}
\label{sec:rq3}
For RQ3, \textbf{\emph{\rqiii}} we aim to analyze if LLMs can understand chains of connections between document and code elements. We present our motivation, approach and results.

\subsection{Motivation}
\label{sec:rq3_motivation}
Building on our earlier findings that LLMs can (i) identify direct
document‑to‑code links (RQ1) and (ii) explain those links (RQ2), we now investigate whether they can also recover \emph{multi‑step chains} that include intermediate artifacts. Documentation often relates to code through chains of connections, where understanding intermediate elements is crucial for maintenance and comprehension. While identifying a single direct link is valuable (e.g., documentation $\rightarrow$ class), real-world software often involves multiple code artifacts that together implement the documented feature. For instance, as discussed in Section~\ref{sec:dataset_creation}, our \textit{Crawl4AI} dataset includes a documentation segment about configuring the \texttt{max\_concurrent\_requests} parameter within the \texttt{Crawler} constructor, which in turn relies on \texttt{DEFAULT\_CONCURRENCY} as a fallback. This scenario forms a trace chain of:
\[
\textit{Documentation} \;\rightarrow\; \texttt{Crawler.\_\_init\_\_}
\;\rightarrow\; \texttt{Crawler.DEFAULT\_CONCURRENCY}
\]
 
Identifying the intermediate artifact reveals the underlying workflow. Therefore, we aim to determine whether LLMs can accurately detect these multi-step connections, which are critical for tasks such as change impact analysis in real-world software systems.

\subsection{Approach}
\label{sec:rq3_approach}
To evaluate the LLMs' ability to identify trace chains, we analyze chain completeness and accuracy. For each trace link identified in RQ1, we analyzed whether the LLMs' could detect intermediate elements between the documentation and the traced code element.

For each generated chain, we analyze by parsing chains into their constituent elements and comparing them against ground truth chains. We evaluate trace chains using a structured classification system:
\begin{itemize}
   \item Complete Match: Chain completely matches the ground truth,
   \item Partial Match (Intermediate): Correct endpoints but different middle elements,
   \item Partial Match (Prefix): Correctly starts with the first endpoint of the chain,
   \item Partial Match (Suffix): Correctly ends with the last endpoint of the chain,
   \item Incorrect: Completely different from ground truth.
\end{itemize}

\subsection{Results}
\label{sec:rq3_results}

\begin{table}
\small
\centering
\caption{Trace Chain Correctness Analysis Across LLMs}
\label{tab:chain_correctness}
\begin{tabular}{@{}llrrrrrr@{}}
\toprule
\textbf{Dataset} & \textbf{Model} & \textbf{Total} & \textbf{Complete} & \textbf{Partial} & \textbf{Partial} & \textbf{Partial} & \textbf{Incorrect} \\
                 &              & \textbf{Chains} & \textbf{Match (\%)}    & \textbf{(Interm.) (\%)} &  \textbf{(Prefix) (\%)} & \textbf{(Suffix) (\%)} & \textbf{(\%)} \\
\midrule
\multirow{3}{*}{\textbf{Crawl4AI}} & Claude 3.5 & 2,200 & 23.0 & 67.8 & 8.3 & 0.0 & 0.0 \\
                          & GPT-4o     & 1,687 & 30.6 & 68.0 & 1.4 & 0.1 & 0.0 \\
                          & o3-mini    & 1,862 & 12.8 & 23.8 & 0.6 & 61.3 & 1.6 \\
\midrule
\multirow{3}{*}{\textbf{Unity C.}}    & Claude 3.5 & 580   & 67.4 & 25.2 & 7.4 & 0.0 & 0.0 \\
                          & GPT-4o     & 412   & 65.8 & 31.6 & 2.7 & 0.0 & 0.0 \\
                          & o3-mini    & 363   & 80.2 & 6.1  & 1.4 & 12.4 & 0.0 \\
\bottomrule
\multicolumn{8}{@{}l}{\small Note: Unity C. is an abbreviation of Unity Catalog} \\
\end{tabular}
\end{table}

Table~\ref{tab:chain_correctness} categorizes the generated trace chains based on their alignment with the ground truth, classifying them as Complete Match, various types of Partial Match (Intermediate, Prefix, Suffix), or Incorrect. The chains considered are the sum across all runs. Two key patterns emerge:

\textit{Errors Rarely Mean Complete Failure.} Across both datasets, outright incorrect chains are almost non-existent, except for o3-mini in the Crawl4AI dataset (1.6\%). When LLMs fail to reproduce exact chains, they typically maintain meaningful connections by preserving either of the endpoints. This demonstrates that even partial traces often contain valuable relationship information.

\textit{Chain Recovery Greatly Varies Across Datasets.} The most striking finding is the substantial difference in complete match rates between the datasets. All LLMs perform substantially better in \textit{Unity Catalog} (structured API documentation) compared to \textit{Crawl4AI} (narrative tutorial-style documentation), with Complete Match scores of 66-80\% vs 13-31\%. This suggests that formal, structured documentation might provide clearer signals for LLMs to accurately trace multi-step connections.

\begin{table}
\small
\centering
\caption{Interior Error Pattern Distribution}
\label{tab:chain_patterns}
\begin{tabular}{@{}llrrr@{}}
\toprule
\textbf{Dataset} & \textbf{Model} &
\textbf{Extended (\%)} & \textbf{Shortened (\%)} & \textbf{Substituted (\%)} \\
\midrule
\multirow{3}{*}{\textbf{Crawl4AI}}
 & Claude 3.5 & 73.4 & 11.3 & 15.2 \\
 & GPT‑4o      & 58.0 & 25.7 & 16.3 \\
 & o3‑mini     & 58.0 & 15.8 & 26.2 \\
\midrule
\multirow{3}{*}{\textbf{Unity Catalog}}
 & Claude 3.5 & 92.5 &  0.0 &  7.5 \\
 & GPT‑4o      & 41.5 & 43.1 & 15.4 \\
 & o3‑mini     & 13.6 & 86.4 &  0.0 \\
\bottomrule
\end{tabular}
\end{table}

\header{Interior Error Patterns.}
In Table~\ref{tab:chain_correctness}, we observed that the cases where the intermediate nodes were incorrect (\emph{Partial‑Intermediate} chains) were the most common type of error. This is expected as we provide the LLM with the starting and end points in the prompt. Table~\ref{tab:chain_patterns} shows the three different patterns of intermediate errors:

\begin{itemize}
  \item \textbf{Extended (extra nodes).}  
        The most frequent pattern overall: 14-93\% of intermediate
        chains have one or more additional artifacts, indicating that
        LLMs often ``over‑connect'' the intended trace chain.
  \item \textbf{Shortened (missing nodes).}  
        In 0-86\% of cases, the LLM compresses the path by omitting
        at least one required step, suggesting an inclination to trim
        what it sees as redundant detail.
  \item \textbf{Substituted (node replaced).}  
        When the error is neither pure addition nor pure omission,
        the LLM swaps an interior node for another, accounting for up
        to 26\% of chains.  Pure re‑ordering without addition/removal
        is negligible (<0.1\%) and is therefore omitted from the Table ~\ref{tab:chain_patterns}.
\end{itemize}

\begin{tcolorbox}[colback=white,colframe=gray!50, coltitle=black,
                  arc=3pt,title=\textbf{Key Findings for RQ3}]
\noindent
\textbf{LLMs do recover multi‑step links, but completeness varies.}  
Exact matches are common on \textit{Unity Catalog} (66-80\%) yet scarce on \textit{Crawl4AI} (13-31\%), indicating that chain recovery greatly varies across datasets.

\medskip
\textbf{When only the interior is wrong, three error patterns dominate.}  
Across all LLMs, interior deviations are mostly \emph{Extended} paths (extra nodes, 14-93\%), followed by \emph{Shortened} paths (missing nodes, 0-86\%), and \emph{Substituted} paths (node swapped, up to 26\%); pure re‑ordering is negligible.

\medskip
\textbf{Endpoints are reliable anchors.}  
Chains that lose both start and end points are below 2\% for every LLM-dataset pair, so even imperfect traces give developers a sound entry and exit for navigation.
\end{tcolorbox}

\section{Discussion}
\label{sec:discussion}

Our three research questions showed that LLMs  
(i) identify document‑to‑code links with high precision but uneven recall (RQ1);  
(ii) explain those links with good, though still partial, semantic fidelity (RQ2); and  
(iii) reconstruct multi‑step chains once the link endpoints are clear (RQ3).  

This section now turns from \emph{what} the LLMs achieve to \emph{how} task-framing influences those results, specifically, how context‑window management (Section \ref{subsec:context-analysis}) and the amount of surrounding documentation (Section \ref{subsec:extended-context}) help or hinder trace discovery. Claude Sonnet 3.5 serves as the running example because it was the strongest LLM overall; when GPT‑4o or o3‑mini behave differently under the same context settings, we highlight those differences in the text.

\subsection{Impact of Context Management Strategies on LLM-Based Traceability}
\label{subsec:context-analysis}

A critical challenge in leveraging LLMs for software traceability is how to effectively present documentation and code for analysis. This challenge stems from two key constraints: the finite context window of LLMs \cite{liu2024lost, wang2024leave} and their tendency to truncate or limit enumerated responses \cite{saito2023verbosity, song2025hansel, nayab2024concise}. Understanding how to manage these constraints is crucial for both the effectiveness and efficiency of using LLMs for traceability.

There are three fundamental matching strategies for structuring traceability queries for LLMs to manage their context, each with distinct advantages and trade-offs:

\header{One-to-one matching strategy.} The most granular matching strategy examines each possible document-code pair individually. For instance, given two documentation segments ($d_1$, $d_2$) and three code artifacts ($c_1$, $c_2$, $c_3$), this strategy would make six separate queries:
\begin{center}
$d_1 \leftrightarrow c_1$, $d_1 \leftrightarrow c_2$, $d_1 \leftrightarrow c_3$\\
$d_2 \leftrightarrow c_1$, $d_2 \leftrightarrow c_2$, $d_2 \leftrightarrow c_3$
\end{center}
While this strategy ensures focused attention on each potential relationship, it requires $N \times M$ API calls for $N$ documents and $M$ code artifacts. The cost multiplies rapidly when including necessary contextual information like directory structures and surrounding code.

\header{One-to-many matching strategy.} This matching strategy examines each documentation segment against all code artifacts simultaneously. Using the same example:
\begin{center}
$d_1 \leftrightarrow \{c_1, c_2, c_3\}$\\
$d_2 \leftrightarrow \{c_1, c_2, c_3\}$
\end{center}
This reduces API calls to $N$ while maintaining focused analysis of each documentation segment. The LLM must identify which of the provided code artifacts, if any, trace to the current document.

\header{Many-to-many matching strategy.} The most computationally efficient matching strategy presents all documentation and code simultaneously:
\begin{center}
$\{d_1, d_2\} \leftrightarrow \{c_1, c_2, c_3\}$
\end{center}

To empirically evaluate these trade-offs, we conducted a controlled experiment comparing our one-to-many matching strategy with a many-to-many matching strategy. We maintained identical conditions across both strategies, including five randomized runs, consistent prompting, and equal access to contextual information. Table~\ref{tab:context-comparison} presents the detailed results of this comparison. We exclude the experiments for one-to-one as it would be super costly.

\begin{table}
\centering
\caption{Impact of Context Management Strategy on Traceability Performance}
\label{tab:context-comparison}
\begin{tabular}{llrrrrrrr}
\hline
\textbf{Dataset} & \textbf{Strategy} & \textbf{API Calls} & \textbf{TP} & \textbf{FP} & \textbf{FN} & \textbf{Prec.} & \textbf{Rec.} & \textbf{F1} \\
\hline
Unity Catalog & One-to-many    & 32  & 116.0  & 17.4  & 39.0   & 0.870 & 0.748 & 0.804 \\
              & Many-to-many   & 1   & 5.0    & 5.2   & 150.0  & 0.496 & 0.032 & 0.061 \\
\hline
Crawl4AI    & One-to-many    & 112 & 440.0  & 22.0  & 207.0  & 0.953 & 0.680 & 0.794 \\
              & Many-to-many   & 1   & 17.6   & 3.2   & 624.4  & 0.829 & 0.027 & 0.053 \\
\hline
\multicolumn{9}{@{}l}{\small Note: Results are averaged over five runs.}
\end{tabular}
\end{table}

The results reveal striking differences in the number of traces identified by each matching strategy. 

Across both datasets, issuing a single many‑to‑many prompt sharply degrades performance: recall falls below 3\% and F1 drops to 5\%, compared with around 80\% under the one‑to‑many strategy. This decline likely stems from output‑token limits \cite{saito2023verbosity, song2025hansel, nayab2024concise}, which prevent the LLM from listing most valid links when all candidates are presented at once. The one‑to‑many strategy avoids that constraint, preserving the high precision reported in RQ1 and most of the recall, at the cost of one request per documentation segment. Practitioners, therefore, face a trade‑off between API cost and trace coverage; with current LLMs, one‑to‑many remains the better choice when accuracy matters.

While requiring only one API call per run, many-to-many matching strategy faces three critical limitations:
\begin{itemize}
\item The combined size of all inputs may saturate or exceed the LLM's context window.
\item Output token limits, typically more restrictive than input limits, can artificially constrain the number of reported traces.
\item Processing everything simultaneously reduces attention to individual document-code relationships, limiting the LLM's ability to focus deeply on specific pairs.
\end{itemize}

To help practitioners balance these trade-offs, we developed cost models for each matching strategy:

\begin{equation}
\label{eq:cost-one-to-one}
C_{\text{one-to-one}} = N \times M \times t \times (i_d + i_a + s + o)
\end{equation}

\begin{equation}
\label{eq:cost-one-to-many}
C_{\text{one-to-many}} = N \times t \times \bigl(i_d + M \times i_a + s + o\bigr)
\end{equation}

\begin{equation}
\label{eq:cost-many-to-many}
C_{\text{many-to-many}} = t \times \bigl(N \times i_d + M \times i_a + s + o_{\text{max}}\bigr)
\end{equation}
Here, $N$ represents the number of documentation segments, $M$ the number of code artifacts, and $t$ the cost per token. Input tokens are broken down into documentation ($i_d$), artifact ($i_a$), and supporting context ($s$) components. The output tokens ($o$) are especially crucial in the many-to-many matching strategy, where $o_{\text{max}}$ represents the LLM's maximum output limit.

Our findings suggest that while the many-to-many matching strategy offers substantial cost savings through reduced API calls, these savings come at the expense of significantly reduced trace identification. The dramatic drop in both true positives and recall indicates that this matching strategy misses many valid traces, due to output limitations and reduced attention to individual relationships \cite{dvivedi2024comparative, khan2022automatic}. For comprehensive traceability analysis, especially in larger projects, the one-to-many matching strategy’s focused examination of each documentation segment provides more reliable and complete results despite its higher computational cost.

\subsection{Impact of Additional Contextual Information on Traceability Performance}
\label{subsec:extended-context}

In RQ1, for each documentation segment sent, we provide the LLM with the list of code artifacts and the directory tree structure of the repository. However, we hypothesize that more context might help the LLM identify additional relevant traces. Here, we explore the impact of a modified approach that provides the entire documentation file (from which a segment was derived) instead of just the documentation segment. To examine the impact of this additional contextual information on the LLM performance, Table~\ref{tab:perf-comp} compares key metrics for the segment-only approach against the file-context approach.

\begin{table}
\centering
\caption{Performance Comparison: Segment-Only vs. File-Context Approaches}
\label{tab:perf-comp}
\begin{tabular}{@{}llrrrrrr@{}}
\toprule
\textbf{Dataset} & \textbf{Approach} & \textbf{TP} & \textbf{FP} & \textbf{FN} & \textbf{Prec.} & \textbf{Rec.} & \textbf{F1} \\
\midrule
\multirow{2}{*}{Crawl4AI} 
  & Segment-Only & 440.0 & 22.0  & 207.0  & 0.953 & 0.680 & 0.794 \\
  & File-Context & 407.8 & 104.2 & 217.6  & 0.796 & 0.652 & 0.717 \\
\midrule
\multirow{2}{*}{Unity Catalog}
  & Segment-Only & 116.0 & 17.4  & 39.0   & 0.870 & 0.748 & 0.804 \\
  & File-Context & 127.4 & 23.4  & 27.0   & 0.845 & 0.825 & 0.835 \\
\bottomrule
\multicolumn{8}{@{}l}{\small Note: Results are averaged over five runs.}
\end{tabular}
\end{table}

As Table~\ref{tab:perf-comp} shows, the file-context approach decreases precision and recall in Crawl4AI. Although it yields a slight increase in recall for Unity Catalog, it still results in more errors (FP + FN) overall. To investigate these outcomes further, we performed a qualitative analysis of false positives. Table~\ref{tab:err-comp} presents the error distributions for both datasets, revealing that Context Boundary Violation (CBV) (which was effectively absent for the segment-only approach) emerges prominently whenever the entire file is provided. In Crawl4AI, CBV constitutes 56.6\% of all error tags under the file-context approach, while in Unity Catalog it accounts for 34.0\%. This suggests that when more context is included, the LLM often traces across segment boundaries and flags extraneous relationships.

\begin{table}
\small
\caption{Error Analysis for Segment‑Only vs.\ File‑Context Approaches}
\label{tab:err-comp}
\begin{tabularx}{\linewidth}{@{}l *{4}{>{\centering\arraybackslash}X}@{}}
\toprule
\multirow{2}{*}{\textbf{Error Group}}
  & \multicolumn{2}{c}{\textbf{Crawl4AI}}
  & \multicolumn{2}{c}{\textbf{Unity Catalog}} \\
\cmidrule(lr){2-3}\cmidrule(lr){4-5}
  & \textbf{Segment\,(\%)} & \textbf{File\,(\%)}
  & \textbf{Segment\,(\%)} & \textbf{File\,(\%)} \\
\midrule
Implicit Assumption Errors (IAE) & 61.3 & 26.3 & 51.9 & 34.0 \\
Phantom Link (PAL)               & 30.7 &  2.0 & 15.5 & 17.7 \\
Architecture Pattern Bias (APB)  &  0.0 &  2.0 & 32.6 & 12.8 \\
Implementation Overlink (IOL)    &  8.0 & 13.2 &  0.0 &  1.4 \\
Context Boundary Violation (CBV) &  0.0 & 56.6 &  0.0 & 34.0 \\
\midrule
\textit{Total Error} &  150 & 152 &  129 & 141 \\
\bottomrule
\end{tabularx}
\end{table}

These error distributions highlight the key difference introduced by file-level context: CBV becomes a major source of false positives, indicating that the LLM’s ability to focus on a specific segment is diluted by the surrounding text. Instead of confining traces within the relevant segment, the LLM ``wanders'' into other parts of the file, triggering incorrect links. So, while including the full documentation file might seem advantageous, it often produces confusion rather than clarity. The additional context allows the LLM to conflate content across segment boundaries, increasing false positives due to CBV. Therefore, practitioners should carefully consider the appropriate level of context when employing LLMs for traceability tasks.

\section{Implications}
\label{sec:implications}
The findings of our paper carry implications for software development \& maintenance practices, for tooling \& automation, and for AI/LLM research and development.

\subsection{Software Development \& Maintenance Practices}
\header{Transforming Development Practices with LLM-Assisted Traceability.}
Our results show that LLMs represent a significant advancement in automated traceability approaches. While existing automated methods like information retrieval (TF-IDF, BM25) and embedding-based techniques (CodeBERT) have helped reduce manual effort, they achieve F1-scores of only 36.2\% to 69.3\% in our experiments. LLMs, reaching up to 80.4\%, offer a new level of accuracy that makes automated traceability more practical for daily development tasks. This improvement matters most in scenarios where traditional automated methods fall short: understanding implicit relationships, interpreting usage examples, and connecting high-level descriptions to specific implementations. For instance, when documentation describes a configuration parameter conceptually, LLMs can identify not just the parameter itself but also the methods that use it and the default values it falls back to. This semantic understanding transforms how teams approach documentation maintenance. During refactoring or API changes, developers can quickly assess the full impact across all documentation, not just places with exact keyword matches. New team members benefit particularly from this capability, as they can explore how concepts mentioned in tutorials connect to actual code implementations, building mental models faster than with traditional search-based approaches. The shift is from keyword-based discovery to semantic understanding, making documentation a more effective resource for both learning and maintenance.

\header{Integrating LLM Capabilities into Team Workflows.}
Successfully incorporating LLMs into development practices requires thoughtful integration that respects both their strengths and limitations. Rather than treating LLMs as a wholesale replacement for existing processes, teams should view them as a new tier in their traceability toolkit. For routine documentation updates and initial impact analysis, LLMs can automate what previously required developer time. However, for critical documentation such as API contracts or security-related descriptions, human review remains essential. The key is establishing clear guidelines about when LLM suggestions can be trusted and when they require verification. Our findings suggest that explicit references and method-level traces are highly reliable, while class-level connections in narrative documentation need more scrutiny. Teams might establish workflows where LLM-generated links are automatically accepted for certain documentation types (like auto-generated API docs) but flagged for review in others (like architectural decision records). This stratified approach maximizes efficiency while maintaining quality. Additionally, teams should document their LLM usage patterns and share successful prompting strategies, building institutional knowledge about how to best leverage these tools. Over time, this creates a feedback loop where developers become better at working with LLMs, and the team develops customized approaches that fit their specific codebase and documentation style.

\subsection{Tooling \& Automation}

\header{Design Principles for LLM-Enhanced Traceability Tools.}
Our experiments show that architectural choices dramatically affect tool performance. The one-to-many matching strategy achieved F1-scores of 79.4\%, while the many-to-many approach dropped to just 5.3\%. This huge difference occurs because LLMs have strict output token limits and struggle to maintain focus when processing everything at once. Similarly, we found that providing entire documentation files actually hurt performance, increasing false positives by 56.6\% as the LLM incorrectly linked content from outside the relevant segment. This challenges the common assumption that more context always helps. Instead, tools need to break documentation into focused, manageable pieces that match what LLMs can effectively process. Based on these findings, future IDE plugins and documentation platforms should preprocess content into appropriate chunks, cache code artifact representations to reduce API costs, and track which documentation sections have been analyzed to prevent the LLM from making connections across segment boundaries. The key is working with LLM limitations rather than against them.

\header{Human-AI Collaboration in Traceability Tasks.}
While LLMs rarely miss fundamental connections (with over 97\% partial accuracy), they achieve full correctness only 42.9\% to 71.1\% of the time. This gap suggests that developers should view LLMs as assistants rather than replacements. The most effective approach combines LLM capabilities with human judgment: LLMs quickly identify potential links and explain basic relationships, while developers verify and refine these suggestions. This requires interfaces that go beyond simple accept/reject buttons. For example, when an LLM correctly identifies the start and end of a trace chain but gets the intermediate steps wrong (which happened in 13\% to 87\% of multi-step chains), developers need tools that let them easily fix the path while keeping the valid parts. Additionally, our results suggest combining LLMs with traditional static analysis. LLMs excel at understanding what documentation means and finding semantic connections, while static analysis can precisely trace execution paths. Tools that blend both approaches could offer developers the best of both worlds: semantic understanding from LLMs and technical precision from static analysis.

\subsection{AI/LLM Research \& Development}

\header{Understanding LLM Limitations in Software Engineering.}
Our analysis uncovered specific patterns in how LLMs fail at software tasks. The most common error type, implicit assumption errors (averaging 39.3\% of all errors), shows that LLMs often confuse their general programming knowledge with the specific codebase they're analyzing. They assume inheritance relationships based solely on class names, expect architectural patterns that do not exist, and create links to methods they think should be there but are not. These are not random mistakes but predictable biases that affect all three LLMs we tested. This suggests the problem lies in how current LLMs process information rather than just needing better training data. Researchers could address these issues by training models to explicitly recognize when not to make connections, developing architectures that better separate general knowledge from context-specific information, or creating prompting techniques that explicitly tell models to avoid assumptions. 

\header{Building Better Evaluation Methods for LLMs in Software Engineering.}
Our study provides both datasets and evaluation methods that future researchers can build on. By examining not just whether LLMs find links but also how well they explain them and trace multi-step connections, we showed that simple accuracy metrics hide important details. A model might correctly identify links but fail to explain them properly, making it less useful for developers who need to understand why connections exist. Our cost analysis, comparing different matching strategies and their API requirements, offers a practical framework for teams deciding how to deploy LLMs within budget constraints. The datasets we created, with 144 documentation segments and 105 code artifacts across two different documentation styles, test both obvious references and subtle implicit connections. Crucially, we used projects created after the LLMs' training cutoff dates to ensure fair evaluation on truly unseen code. Future work can extend this approach by testing on larger codebases, exploring whether specialized training improves performance, or developing standardized benchmarks for software engineering tasks. As LLMs become more common in development tools, we need rigorous evaluation methods that capture both their strengths and limitations in real-world scenarios.

\section{Threats to Validity}
\label{sec:threats}
We discuss threats to the validity of our study across four dimensions: internal, external, construct, and conclusion validity.

\subsection{Internal Validity}

A key threat to internal validity arises from our ground-truth dataset creation. While we followed a systematic, step-by-step process for labeling and verifying trace links, there remains some inherent subjectivity when identifying implicit relationships between documentation and code. In particular, deciding whether a segment ``indirectly references'' a code element can vary between reviewers. We mitigated this by providing clear, explicit guidelines and holding iterative discussions in cases of disagreement.

Another threat is the non-deterministic nature of large language models (LLMs). Even with standardized prompts and carefully controlled parameters, repeated runs may yield variations in outputs. To mitigate this, we conducted \emph{five runs} for each experiment and each LLM, using different random seeds and shuffled input orders, then aggregated the outcomes to minimize outlier effects.

Since we evaluated multiple LLMs, each with distinct pretraining corpora, subtle differences in prompt engineering could influence outcomes. Minor variations in how context is segmented or how instructions are phrased may alter performance. To reduce these risks, we applied a consistent prompting format (the one-to-many approach) across all LLMs, caching shared content (e.g., code blocks or file listings) so that each model received the same inputs.

Finally, although we selected repositories created after the LLMs’ training cutoff dates to reduce data contamination, there remains a remote possibility that some portion of these codebases—or their broader patterns-were known to the LLMs. We minimized this risk by verifying that these repositories were indeed \emph{newer projects}, and by confirming that pivotal code elements had not existed in older forks or widely shared archives.

\subsection{External Validity}

Our study focuses on two open-source projects created after the LLM training cutoffs. This methodological choice helps ensure that none of our evaluated LLMs had been trained on the exact project code or documentation. These newer repositories also have varied documentation styles, including tutorial-style sections and API references, allowing us to evaluate different documentation formats. Additional studies on a wider range of systems (e.g., closed-source legacy or industrial codebases) could further test how these methods generalize, especially if different documentation styles or domain constraints are involved. Likewise, future work may compare an even broader set of LLMs, as emerging models continue to evolve rapidly.

\subsection{Construct Validity}

Construct validity threats involve how we define and operationalize ``trace link'', ``relationship explanation'', and ``trace chain''. Our criteria distinguish between explicit references, implicit references, and multi-step connections. While these definitions are grounded in prior literature, other researchers might categorize or label relationships differently. We mitigated this through pilot testing on sample segments and ensuring our labeling guidelines were sufficiently detailed to capture both straightforward and subtle forms of traceability.

Our evaluation metrics---precision, recall, F1-scores for link detection, and correctness/completeness for relationship explanations---reflect commonly accepted measures in traceability. Yet, developers might additionally value metrics such as usability or domain relevance. We partially accounted for this by examining detailed explanation quality (e.g., completeness of reasoning) rather than just binary correctness. However, more nuanced or domain-specific constructs (e.g., compliance-level trace explanations) could require specialized measurement strategies.

\subsection{Conclusion Validity}

One potential threat to conclusion validity is the inherent variability of LLM-based systems. Providers routinely update models, and performance on identical prompts may drift over time. We mitigated this by recording the exact version identifier for each LLM used in our experiments, carrying out all runs within a defined time window. This ensures that subsequent researchers could reference the same model versions should they wish to replicate our procedures.

We further strengthened our conclusions through repeated experiments (five runs per LLM for all experiments) and standardized prompt structures, thus capturing a more robust view of performance. These measures reduce the likelihood that any single run’s outlier behavior skews the findings. While future model updates or additional hyperparameter tuning might alter results, our detailed documentation of methods and model versions supports replicability.

Lastly, we acknowledge that the scope of our findings is bounded by the specific LLMs, prompt structures, and projects we evaluated. Although we selected \emph{new} open-source repositories and performed a comprehensive analysis of documentation-to-code links, different codebases or LLM architectures might reveal additional nuances. Our work nevertheless provides a reproducible foundation for subsequent studies to expand upon or refine.

\section{Conclusion}
\label{sec:conclusion}
This work set out to systematically evaluate how LLMs perform on documentation-to-code traceability tasks, using two newer open-source projects specifically chosen to reduce the risk of data contamination from the LLMs’ pretraining corpora. Overall, our results indicate that LLMs substantially outperform traditional information-retrieval baselines when identifying direct trace links, yet offer only partial completeness in their explanatory details. In multi-step or chain-based relationships, the LLMs consistently achieve high endpoint accuracy but sometimes struggle with precisely identifying intermediate elements. A closer look at these results reveals three themes. First, LLM-based trace discovery exhibits decent performance, providing clear evidence of their ability to interpret documentation. Second, while LLMs seldom fail entirely on explanation tasks---often capturing the core relationship---they remain prone to omissions of more subtle yet potentially important nuances.

These results indicate that LLMs have strong potential for significantly reducing the manual effort required to create trace links, especially when documentation follows coherent, well-structured conventions that make it easier for a language model to associate relevant code artifacts. At the same time, developers should remain aware that LLM-generated explanations may omit subtle but important information. Our findings thus point to a collaboration model where LLMs provide an efficient ``first pass'' at trace creation and refinement, while human reviewers handle complex or domain-specific nuances. In practice, this suggests integrating simple accept/reject controls for LLM‑suggested links into IDEs or documentation portals, so teams gain speed without sacrificing accuracy.

\bibliographystyle{ACM-Reference-Format}
\bibliography{references}

\appendix
\label{sec:appendix}
\section{Appendix}
\label{appendix:supplementary}

Below we provide additional details of our methodology. Figure~\ref{fig:prompt-structure} in the main body illustrates our prompt structure for documentation-to-code traceability. Here, we also include the code for asynchronous document processing used in our experiments.

\begin{lstlisting}[language=, caption={Full Evaluation Prompt}, label={lst:full_evaluation_prompt}]
async def process_document(idx: int, request: Dict, total: int, system_message: str, results_dir: str) -> List[Dict]:
    """
    Process a single document request asynchronously.
    Prints status messages, builds the prompt, calls the LLM concurrently,
    and processes the response.
    """
    print(f"\nProcessing document {idx}/{total}")

    # Prepare cached content with all available artifacts and optional document context
    cached_content = {
        "available_artifacts": [
            {
                "artifact_id": artifact["artifact_id"],
                "title": artifact["title"],
                "location": artifact["location"],
                "content": artifact["content"]
            }
            for artifact in request["artifact_list"]
        ],
        "directory_tree": request["directory_tree"]
    }

    if request.get("document_context"):
        cached_content["document_context"] = request["document_context"]

    cached_message = json.dumps(cached_content, indent=2)

    # Construct prompt
    prompt_dict = {
        "task": "Documentation to Code Traceability Analysis",
        "description": "Analyze the documentation snippet and identify code artifacts that implement or relate to the documented functionality. Use only artifact names from the provided list.",
        "data": {
            "document": {
                "text": {
                    "value": request["doc_text"],
                    "description": "Documentation text snippet to analyze for trace links"
                },
                "location": {
                    "value": request["doc_location"],
                    "description": "Location information for context"
                },
                "file": {
                    "value": "",  # request["document_file"] omitted to run without file
                    "description": "File from which the document.text snippet was extracted, for context, the file may or may not be included"
                }
            }
        },
        "instructions": {
            "steps": [
                "1. Focus ONLY on the provided 'text' field when identifying traces",
                "2. The location and file information are provided for context only - DO NOT trace to elements mentioned elsewhere in the file",
                "3. Analyze the specific text snippet to identify explicitly mentioned code elements",
                "4. Identify any code elements demonstrated in usage examples within this text",
                "5. For each identified artifact:",
                "   - If it's explicitly mentioned in the text snippet, mark as 'explicit'",
                "   - If it appears in usage examples within this text but isn't directly discussed, mark as 'implicit'",
                "   - Explain the specific nature of how it relates to the documentation",
                "   - Identify if it's part of a chain/pathway to other artifacts",
                "6. Only trace to:",
                "   - Classes and their usage",
                "   - Methods and their usage",
                "   - Class-level attributes that form part of the public interface and their usage",
                "7. DO NOT trace to:",
                "   - Individual statements within methods",
                "   - Implementation details",
                "   - Local variables or parameters",
                "   - Elements mentioned outside the provided text snippet",
                "8. For each relationship:",
                "   - Provide specific evidence from the text",
                "   - Explain the nature of the relationship in detail",
                "   - Identify any intermediate steps or dependencies",
                "9. Make sure to include all required fields in your response",
                "10. Ensure you only trace to the included list of code snippets returning only titles within that list given",
                "11. For any traced class, also trace to its base classes and implementing classes if they are in the available artifacts list",
                "12. For traceability pathways:",
                "   - Always start with the document name (extracted from location)",
                "   - Use exact artifact titles for code elements",
                "   - Strictly follow 'A -> B -> C' format with spaces around arrows",
                "   - Must end at the traced artifact",
                "   - Explain why each intermediate step is necessary"
            ],
            "output_format": {
                "artifacts": [
                    {
                        "artifact_id": {
                            "type": "integer",
                            "description": "Unique identifier for the artifact"
                        },
                        "title": {
                            "type": "string",
                            "description": "Exact artifact title from available_artifacts"
                        },
                        "relationship": {
                            "type": "string",
                            "enum": ["explicit", "implicit"],
                            "description": "Whether artifact is explicitly mentioned or implicitly used"
                        },
                        "relationship_type": {
                            "type": "string",
                            "description": "Nature of relationship (implements, extends, uses, etc.)"
                        },
                        "relationship_explanation": {
                            "type": "string",
                            "description": "Detailed explanation with evidence from text"
                        },
                        "trace_chain": {
                            "type": "string",
                            "description": "Format: doc_name.md -> Artifact1 -> Artifact2"
                        },
                        "trace_chain_explanation": {
                            "type": "string",
                            "description": "Explanation of chain relationships"
                        }
                    }
                ]
            }
        }
    }
\end{lstlisting}

\begin{lstlisting}[language=, caption={Prompt for LLM as a Judge in RQ2}, label={lst:llm_judge_prompt}]
prompt = {
            "task": "Judge Relationship Description (3-category)",
            "context": {
                "code": code_content,
                "documentation": doc_text
            },
            "descriptions": {
                "predicted": predicted,
                "ground_truth": ground_truth
            },
            "instructions": """
            You are an expert judge comparing two relationship descriptions:
            (1) predicted vs. (2) ground_truth.

            Focus on whether the predicted description gets the high-level 
            relationship right. Do NOT penalize minor omissions 
            about code details or parameters. Only label as partially_correct if 
            the predicted text contradicts or significantly misunderstands a 
            major aspect of the ground truth. Everything else can be correct.

            Label as:
            - "correct" if the predicted description basically captures the 
                same high-level relationship described by the ground truth, 
                even if some minor details are omitted.
            - "partially_correct" if it covers some of the core idea but 
                introduces a misunderstanding or omits a crucial aspect that 
                changes the overall meaning.
            - "incorrect" if it is largely or completely inconsistent with 
                the ground truth (e.g., describing a different class relationship 
                or functionality).

            Provide:
            - alignment_label: correct, partially_correct, or incorrect
            - justification: short reason
            - error_type: if not correct (e.g., "major_contradiction")

            IMPORTANT:
            - Do NOT generate new content; only evaluate alignment. 
            - Treat missing details or code specifics as unimportant, 
                as long as the main relationship is correct.
            """,
            "output_format": {
                "alignment_label": "string", 
                "justification": "string",
                "error_type": "string"
            }
        }

        system_message = (
            "You are an impartial evaluator. Determine whether the predicted relationship "
            "is correct, partially_correct, or incorrect relative to the ground_truth. "
            "Do not generate a new explanation. Ensure your response is in JSON only."
        )
\end{lstlisting}

\end{document}